\documentclass[12pt]{article}
\usepackage{amssymb,amsmath,amsfonts,geometry,graphicx,color,setspace,natbib,hyperref}
\usepackage{kpfonts}
\usepackage[tableposition=top, labelfont=bf]{caption}
\usepackage{booktabs}
\usepackage[flushleft]{threeparttable}
\usepackage{subcaption}
\usepackage{siunitx}
\usepackage[table,xcdraw]{xcolor}
\usepackage{placeins}
\usepackage{lineno}
\usepackage{natbib}
\usepackage{float}
\usepackage[T1]{fontenc}
\usepackage{inconsolata}

\usepackage{tabularx} 
\usepackage{dsfont}
\usepackage[nameinlink,capitalise]{cleveref}

\DeclareCaptionFont{xipt}{\bfseries\fontfamily{qag}\selectfont}
\usepackage[font=xipt, labelfont=bf, textfont=normalfont]{caption}

\usepackage{titlesec}
\titleformat*{\section}{\fontfamily{phv}\selectfont\bfseries\LARGE}
\titleformat*{\subsection}{\fontfamily{phv}\selectfont\bfseries\large}
\titleformat*{\subsubsection}{\fontfamily{phv}\selectfont\bfseries\normalsize}

\usepackage{color, colortbl}
\definecolor{Gray}{gray}{0.9}
\definecolor{cerulean}{rgb}{0.0, 0.48, 0.65}
\definecolor{vermilion}{rgb}{0.89, 0.26, 0.2}
\hypersetup{
    colorlinks,
    linkcolor={cerulean},
    citecolor={cerulean},
    urlcolor={cerulean}
}
\bibpunct{\textcolor{cerulean}{(}}{\textcolor{cerulean}{)}}{,}{a}{}{;}

\makeatletter
\AtBeginDocument{
\renewcommand\NAT@open{\color{cerulean}(}}
\makeatother

\usepackage{abstract}

\usepackage[symbol]{footmisc}
\renewcommand{\thefootnote}{\fnsymbol{footnote}}

\usepackage{siunitx}
\DeclareSIUnit{\ugpcm}{\micro\gram\per\cubic\meter}

\begin{document}




\begin{titlepage}
\vspace*{-2.5cm}
\begin{flushleft}
\LARGE\fontfamily{qag}\selectfont\textbf{Estimating the Local Air Pollution Impacts of Maritime Traffic: A Principled Approach for Observational Data\footnote[2]{We thank the Marseille Port officials for sharing their port call data, Méteo-France for providing us their weather data, AtmoSud for openly sharing their air pollution data on their website, and Milena Suarez-Castillo for providing road traffic data from the DIR Méditerranée. We are grateful to Tirthankar Dasgupta for his guidance on the computation of Fisherian intervals and to Stéphane Shao for developing the matching algorithm. We thank Michela Baccini, Sylvain Chabé-Ferret, Olivier Chanel, Augustin Colette, Clément de Chaisemartin, Tatyana Deryugina, Francois Libois, Mirabelle Muûls, Hélène Ollivier, Thomas Piketty, Laure de Preux, Quentin Lippmann, Philippe Quirion, Adam Rosenberg, Thiago Scarelli, Katheline Schubert, Georgia Thebault, Ulrich Wagner, as well as participants to the PSE Applied Economics, PSIPSE and REM seminars and conference participants from EAERE and FAERE for their feedback. Léo Zabrocki and Marion Leroutier acknowledge the support of the EUR grant ANR-17-EURE-0001.}}
\end{flushleft}

\renewcommand*{\thefootnote}{\arabic{footnote}}

\begin{flushleft}
\large\fontfamily{qag}\selectfont{
Léo Zabrocki\footnote{RFF-CMCC European Institute on Economics and the Environment (EIEE), Milan, Italy. Email: \url{leo.zabrocki@gmail.com}. Corresponding author. }\\[2pt]
Marion Leroutier\footnote{Misum, Stockholm School of Economics, Stockholm, Sweden. Email: \url{marion.leroutier@hhs.se}}\\[2pt] 
Marie-Abèle Bind\footnote{Biostatistics Center, Massachusetts General Hospital, Boston, MA, USA. Email: \url{ma.bind@mail.harvard.edu}}}
\\ [0.3cm] 
\end{flushleft}

\begin{abstract}
\small	
\noindent We propose a new approach to estimate the causal effects of maritime traffic when natural or policy experiments are not available. We apply this method to the case of Marseille, a large Mediterranean port city, where air pollution emitted by cruise vessels is a growing concern. Using a recent matching algorithm designed for time series data, we create hypothetical randomized experiments to estimate the change in local air pollution caused by a short-term increase in cruise traffic. We then rely on randomization inference to compute nonparametric 95\% uncertainty intervals. We find that cruise vessels' arrivals have large impacts on city-level hourly concentrations of nitrogen dioxide, particulate matter, and sulfur dioxide. At the daily level, road traffic seems however to have a much larger impact than cruise traffic. Our procedure also helps assess in a transparent manner the identification challenges specific to this type of high-frequency time series data.

\end{abstract}
\vspace{0.3cm}
{\fontfamily{qag}\selectfont \textbf{Website:}} \url{https://lzabrocki.github.io/cruise_air_pollution/}\\[3pt] 
{\fontfamily{qag}\selectfont \textbf{Replication Data:}} \url{https://osf.io/v8aps/}
\vspace{0.3cm}
\setcounter{page}{0}
\thispagestyle{empty}
\end{titlepage}
\pagebreak \newpage

\doublespacing
\setcounter{footnote}{0} 
\renewcommand*{\thefootnote}{\arabic{footnote}}




\section{Introduction}

Maritime traffic generates economic benefits but also comes with environmental and health costs. Particulate matter pollution induced by maritime traffic was estimated to cause 60,000 premature deaths worldwide in 2007, the highest burden being born by the Mediterranean area \citep{corbett2007mortality}. In that region, local environmental organizations and media have recently raised increasing concerns over the potential health impacts due to air pollution emitted by cruise vessels \citep{Friedrich2017, Chrisafis2018}. Estimating the impact of maritime traffic on local air pollutant concentrations is however empirically challenging. Complex meteorological patterns prevail along coastal sites and influence the dispersion of air pollutants. Besides, ports are often located near major roads, making it difficult to disentangle the specific contribution of vessels to local air pollution. While atmospheric scientists can evaluate the proportion of an air pollutant concentration originating from vessel traffic, their methods are difficult to combine with counterfactual analysis preferred by economists. At the same time, in the absence of natural experiments or regulatory changes, quasi-experimental research designs are of little help to assess the effects of maritime traffic on city-level air pollution.

In this study, we propose a new approach to measure the causal effects of vessel traffic on air pollution by creating hypothetical experiments from high-frequency time series data. To illustrate our method, we focus on cruise vessel traffic in Marseille, France, which is the fourth largest Mediterranean port for cruise vessels in 2019, and a relatively polluted city by European standards. We start by combining time series data on cruise traffic, weather parameters, and air pollutant concentrations over the 2008-2018 period. By leveraging on variation in vessel traffic, we try to emulate hypothetical randomized experiments targeted for estimating the impact of a short-term increase in cruise traffic on air pollutants. To better capture the temporal chemistry of air pollutants reaction, we carry out two analyses: one at the hourly level and one at the daily level. We define treatment as hours or days with cruise vessels entering the port. Using a recent constrained pair-matching algorithm designed for time series data \citep{sommer2018comparing}, we construct similar pairs of treated and control time series. Contrary to propensity score matching, the algorithm we use is particularly suited to balance the lags of covariates within matched pairs, which is essential when working with time series.  Once pairs are matched, we assume that the increase in cruise traffic is as-if randomized conditional on a set of observed weather parameters and calendar indicators.

Compared to an approach based on a multivariate regression model, matching has several advantages in our setting. First, it adjusts nonparametrically for observed covariates known to affect pollution in a non-linear way, such as weather parameters. Second, it helps better evaluate the initial imbalance across treated and control units: as cruise traffic has a strong seasonality, it is important to prune control units which do not belong to the common support of the data to avoid model extrapolation \citep{king2006dangers, ho2007matching, stuart2010matching, imbens2015matching}. Third, matching is more transparent than a regression approach to understand which observations are used as counterfactuals for treated units \citep{rosenbaum2010design, rosenbaum2018observation}. 

Since our matching procedure drastically reduces the initial sample size, our preferred inference approach relies on randomization-based inference. It does not rely on large-sample approximation nor makes assumptions on the distribution of the test statistic of interest \citep{fisher1937design, rubin1991practical, ho2006randomization, rosenbaum2010design, imbens2015causal, dasgupta_rubin_2021}. Concretely, to quantify the uncertainty around our estimates, we build 95\% Fisherian intervals that give the range of constant effects supported by the data. This mode of inference relies on the assumption of a constant unit-level treatment effect, which may not hold in our context. We therefore also quantify uncertainty using Neyman's mode of inference \citep{neyman1923applications}, which focuses on average treatment effects, and the recent approach developed by \cite{wu2021randomization} to make randomization inference asymptotically conservative when effects are heterogeneous.

\begin{figure}[!ht]
    \caption{Regularity in Daily Cruise Vessel Traffic over the 2008-2018 Period.}
    \centering\includegraphics[width=\linewidth]{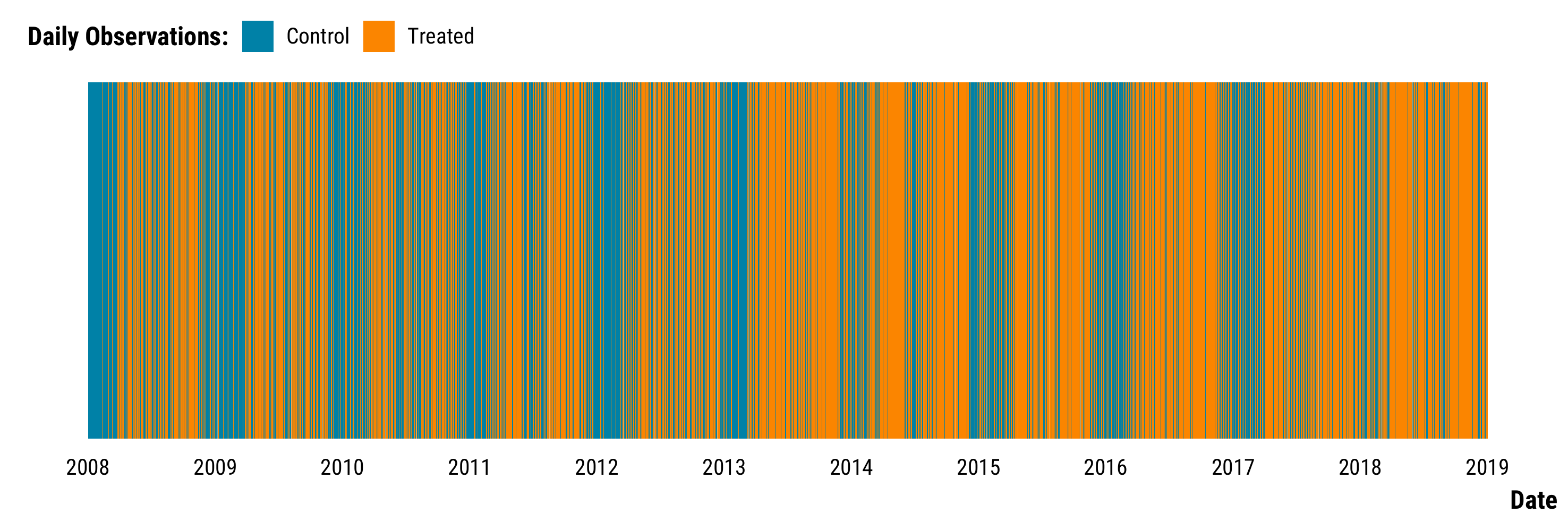}
\label{fig:regularity}
\caption*{\textit{Notes:} Each blue line is a day without cruise traffic and each orange line is a day with cruise traffic.}
\end{figure}

Our transparent approach highlights the challenge of finding, within the time series, a suitable counterfactual for our treated units. Only 4\% of treated units are matched to similar control units at the hourly level, and 8\% at the daily level. This is due to fact that cruise vessel is very regular, as shown on \Cref{fig:regularity}. Besides, up to 20\% of matched pairs are temporally close from each other and might suffer from spillover effects. Despite these limits, our matching procedure is successful in creating well-balanced pairs of treated and control units.

At the hourly level, we find that the arrival of cruise vessel could increase nitrogen dioxide (NO$_{2}$) concentrations between 5\% and 25\%, coarse particulate matter (PM$_{10}$) between 3\% and 27\%, and sulfur dioxide (SO$_{2}$) between 4\% and 109\%. Ozone concentrations (O$_{3}$) simultaneously could decrease up to 14\%, which could be consistent with the titration of this air pollutant due to an increase in nitrogen oxide \citep{diesch2013investigation, eckhardt2013influence, merico2016influence}. Contrary to hourly level results, we do not observe any clear impacts of cruise traffic on air pollution at the daily level: point estimates are null but imprecise. This lack of signal at the daily level is however consistent with a measurement campaign carried out by the local air quality monitoring agency in Marseille \citep{atmosud_quelle_2019}. Besides, the influence of road traffic on air pollutants emitted to a large extent by cars---such as NO$_{2}$---is more discernible in the data. The higher salience of plumes emitted by cruise vessels and the potential larger concerns over this source of air pollution is an interesting area for future research. We complete our main results with an extensive set of robustness checks on issues related to unmeasured confounding, outliers, missing data and low statistical power.

Our paper contributes to the literature in several ways. First, our method is an alternative to atmospheric science methods examining contribution to air pollution by sectors. Atmospheric scientists have relied on two complementary approaches to estimate the contribution of vessel traffic to city-level pollution \citep{mueller2011ships, apice2013, viana2014impact, liu_health_2016, merico2016influence, atmosudbis2018, murena2018impact, liu_emissions_2019, sorte2020impact}. The first approach is model-based. It starts by establishing an emission inventory based on vessels' characteristics and activities. It then infers how emissions turn into concentrations using a dispersion model. The validity of this approach depends on the quality of the emission inventory and the validity of the dispersion model. The second approach is based on source apportionment methods, which require dedicated measurement campaigns with sensors deployed in the city at different seasons. The samples are then analyzed in the laboratory to detect chemical signatures and trace back the likely origin of the particles. The main limits of this second approach is that it cannot be used for gaseous pollutants and the measurement campaigns are conducted over a short period of time. Compared to these two approaches, our method tries to directly relate---without any modeling or need to analyze samples---the variation in maritime traffic to changes in the air pollutant concentrations observed at monitoring stations.

Second, our study also participates to recent works in economics examining the impact of vessel traffic on air pollution and its subsequent adverse health effects with observational data \citep{moretti2011pollution, zhu2021effects, hansen2022uncharted, klotz2022local}. Several of these studies analyze the effects of fuel content regulation on air pollution, which is a convincing source of identification that can be analyzed with time-series regression discontinuity and a difference-in-difference strategies. However, many port cities around the world do not belong to emission control areas and such research designs cannot be implemented to inform future regulation policies. We instead exploit the available variation in vessels traffic to emulate hypothetical experiments. Our approach should be widely applicable to other contexts because observational data on weather, air pollution, and port call statistics are easy to access in several port cities and over a long period of time. 

Third, our approach could be used as a template to help strengthen the design and analysis stages of the growing literature exploiting exogenous transport shocks to estimate the acute health effects of air pollution \citep{moretti2011pollution, schlenker2016airports, knittel2016caution, bauernschuster2017labor, zhong2017traffic, simeonova2021congestion, godzinski2019short, giaccherini2021particulate}. Matching and randomization inference have already proven to be beneficial in air pollution studies on health outcomes that do not
rely on natural experiments \citep{baccini2017assessing, forastiere2020assessing, sommer2021assessing, lee2021discovering}.  Even when we exploit credible source of identification, revealing the common support of the data and laying out the mode of inference is recommended in the statistics literature \citep{rubin1991practical, gutman2012analyses, zigler2014point, bind2019bridging, bind2019causal}. We clearly show in this study how matching helps evaluate the part of the data for which we can draw our inference upon since covariates balance is achieved without a parametric model. Once similar treated and control units are matched, we find more intuitive to exploit the hypothetical treatment allocation as the source of variation rather than assuming, as it is often the case, that observations have been sampled from a larger population \citep{abadie2020sampling}. If randomization inference has recently been the subject of a renewed interest in social sciences \citep{ho2006randomization, cohen2010free, bowers2011fisher, gerber2012field, athey2017econometrics, hess2017randomization, bowers2020causality} and statistics \citep{cattaneo2015randomization, ding2016randomization, keele2019randomization, mackinnon2020randomization, caughey2021randomization, wu2021randomization, zhao2021covariate}, it is comparatively rather underused in environmental economics. We make great efforts to clearly explain the advantages and drawbacks of this mode of inference but also how to concretely implement it. Annotated codes and supplementary materials are available on this \href{https://lzabrocki.github.io/cruise_air_pollution/}{website}. Our data are archived on a \href{https://osf.io/v8aps/}{Open Science Framework} repository.

The rest of our paper is organized as follows. In sections \ref{sec:data} and \ref{sec:research_design}, we present our data and describe the research design we rely on. In section \ref{sec:results}, we present the results and their robustness checks. In section \ref{sec:discussion}, we discuss the advantages but also the limits of our approach and reflect on promising paths for future research.




\section{Data}\label{sec:data}

\begin{figure}
\caption{Map of Marseille's Port surrounding Area and Hourly Vessel Traffic Variation.}
\begin{subfigure}{0.9\textwidth}
    \caption{}
    \centering\includegraphics[scale=0.8]{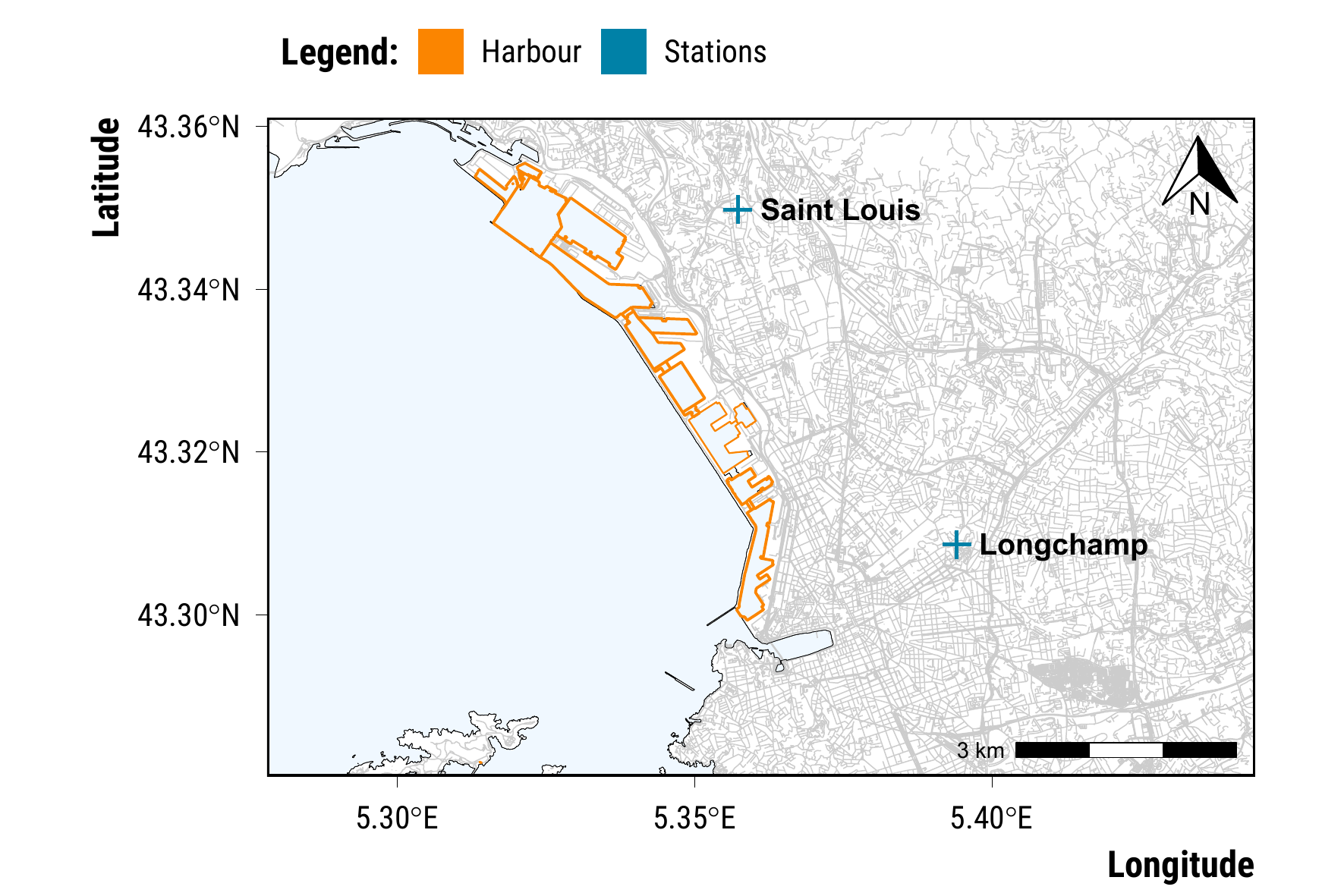}
\label{fig:sub1}
\end{subfigure} \\[1ex]
\begin{subfigure}{0.9\textwidth}
    \caption{}
    \centering\includegraphics[width=\textwidth]{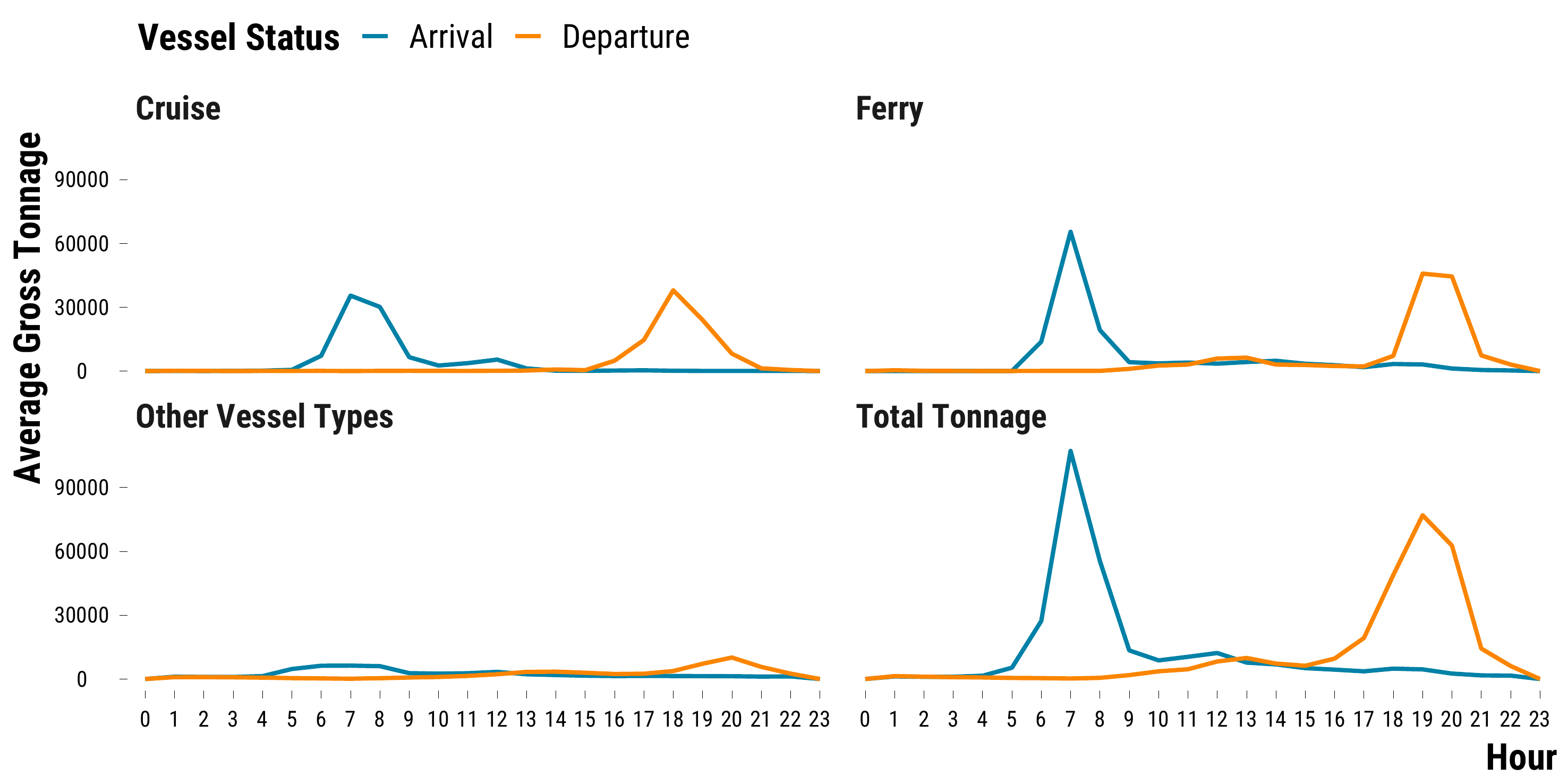}
\label{fig:sub3}
\end{subfigure}
\label{fig:data}
\caption*{\textit{Notes:} Panel A displays a map of Marseille city with its port and the two air quality monitoring stations located in Lonchamp and Saint-Louis neighborhoods. Grey lines represent the road network of the city. Panel B shows the average hourly variation in the gross tonnage of vessel arriving and departing the port. Gross tonnage is a unitless measure of the volume of a ship.}
\end{figure}

We built two datasets for the 2008-2018 period, one at the hourly level with 96,432 observations, and one at the daily level with 4,018 observations. Below we detail the data sources and variables used. In the Data section of our \href{https://lzabrocki.github.io/cruise_air_pollution/}{website}, we report additional information on the data wrangling procedure and carry out a full exploratory data analysis \citep{tukey1977exploratory, tufte1985visual, cleveland1993visualizing}.

\subsection{Vessel Traffic Data}

Most of the maritime passenger traffic occurs in the part of the port located in the city centre of Marseille. This means that a large fraction of the city's residents is potentially exposed to vessels' emissions. This is not an unusual setting as many Mediterranean cities have a port located in the city centre due the development of their economic activities over the centuries. Furthermore, in the absence, until recently, of stringent standards on the pollution content of maritime fuels, cruise vessels entering Marseille port were only subject to the standards imposed to passengers vessels in the European Union. These standards impose a sulfur content of fuels that remains fifteen times higher than the standard in Emission Control Areas existing in Northern Europe and North America.  

We obtained data on 41,015 port calls from the Marseille Port authority. They represent the universe of all port calls between 2008 and 2018. For each vessel docking at the port, we know the exact date and hour of arrival and departure, as well as its name, its type, and its gross tonnage, which is a nonlinear and unitless measure of a vessel's overall internal volume. This measure of a vessel's volume can be related to its emissions of air pollutants and has been used in other studies as a proxy for the intensity of vessel traffic \citep{contini2011direct, moretti2011pollution}. Using information on vessel characteristics, we defined three broad categories: cruise, ferry, and other types of ships. We then calculated, for each vessel type, the total number of vessels and the sum of gross tonnage entering and leaving the port at the hourly and daily levels. As shown in the Panel B of \Cref{fig:data}, vessel traffic is regular: most vessels dock in the port in the morning and leave in the evening. 

\subsection{Air Pollution and Weather Data}

We retrieved air pollution data from the two background monitoring stations managed by AtmoSud, the local air quality agency. The first station, Saint-Louis, is the closest to the cruise terminal. It is located two kilometers away from the cruise terminal (North-Western extremity of the port) and six kilometers away from the ferry terminal (South-Eastern extremity of the port) (See Panel A of \Cref{fig:data}). It only monitors NO$_{2}$ and PM$_{10}$. The second station, Longchamp, is located six kilometers away from the cruise terminal and three kilometers away from the ferry terminal (See Panel A of \Cref{fig:data}). The Longchamp station monitors NO$_{2}$, SO$_{2}$, ozone (O$_{3}$), PM$_{2.5}$ and PM$_{10}$. Sulphur oxides (SOx), nitrogen oxides (NOx), and fine particulate matter are emitted to the atmosphere as a direct result of the combustion of maritime fuel \citep{sorte2020impact}. SOx and NOx emissions directly produce NO$_{2}$ and SO$_{2}$, and contribute to the formation of secondary pollutants such as particulate matter of a larger size (i.e., PM$_{2.5}$ and PM$_{10}$), and O$_{3}$ \citep{viana2014impact}.

Weather data come from Météo-France, the French national meteorological service. We obtained data from the closest weather station, located 25 kilometers away from the city center, at Marseille airport. We calculated hourly and daily values for weather variables: rainfall height (mm), average temperature (°C), humidity (\%), wind speed (m/s), and wind direction measured on a 360 degrees compass rose where 0° is North.

To avoid losing statistical power, we imputed missing values of cruise gross tonnage, air pollutant concentration and weather parameters. We relied on the chained random forest algorithm provided by the \textbf{\textsf{R}} package missRanger \citep{missranger}. There were no clear missingness patterns for these variables and we checked with simulation exercises that this algorithm had a relatively good performance for imputing missing values, even if it could sometimes result in large discrepancies.

\subsection{Road Traffic Data}

We obtained hourly data on the average flow of vehicles and road occupancy rates over the 2011-2016 period from the \textit{Direction Interdépartementale des Routes}, a decentralized state administration in charge of managing, maintaining, and operating roads. We selected hourly data for the six traffic monitoring stations with the best available recordings, two located North and four located East of the city. As measures of road traffic, we focus on the hourly flow of vehicles (number of vehicles) and the occupation rate of the road (\%).




\section{Research Design}\label{sec:research_design}

We conceptualize plausible but hypothetical randomized experiments to estimate the short-term effects of an increase in vessel traffic on air pollutant concentrations in Marseille. We follow a causal inference pipeline conceived to analyze observational data in a rigorous and transparent manner \citep{rubin2008objective, rosenbaum2010design, bind2019bridging, sommer2021assessing}.

\subsection{Stage 1: Formulating Plausible Interventions on Vessel Traffic}

We are interested in the following causal question: \textit{Does cruise vessel traffic contribute to background air pollutant concentrations in Marseille?} The ``ideal" experiment would randomly allocate hours or days to high versus low cruise vessel traffic. We could then confidently attribute the resulting differences in pollutant concentrations to vessel emissions. In the absence of such randomized experiment in Marseille, we try to approximate an experimental setting by comparing pairs of short time series that are as similar as possible on a set of \textit{observed} covariates but differ in their level of vessel traffic. We define below our hypothetical randomized experiments using the framework of the Neyman-Rubin Causal Model \citep{rubin1974estimating, holland1986statistics, rubin2005causal}. We conceive two hypothetical experiments: one experiment at the hourly level to test if an increase in cruise traffic affects hourly air pollutant concentrations in the very short-run; and one experiment at the daily level to examine if an increase in cruise traffic affects daily average concentrations.

The units, which we index by \textit{t} (\textit{t} = 1, \ldots, T), are either hours or days spanning over the 2008-2018 period, depending on the time scale of the experiment considered. At the hourly level, $V_{t}$ is the sum of the gross tonnage of cruise vessels docking in the port during hour \textit{t}. We focus on the pollution impact of cruise vessels arriving in the port rather than aggregating arrivals and departures, because the pollution impact of traffic is likely to depend on the direction of the flow. For example, cruise vessels entering the port may take time to finish maneuvering and generate emissions while they are docked. In contrast, cruise vessels leaving the port may start running their engines a few hours before effectively leaving, and therefore generate pollution over a long period of time. Here, we focus on cruise vessels' arrivals. Our treatment indicator is $W_{t}$ and takes two values:

\begin{equation}\label{eq:1}
W_{t} =
    \begin{cases}
      1 & \text{if $V_{t}>0$}\\
      0 & \text{if $V_{t}$=0}
    \end{cases}       
\end{equation}

\noindent Hourly units with $W_{t}$ equal to one are considered as ``treated" while units with $W_{t}$ equal to zero belong to the control group. A treated hour is an hour with some cruise vessel arrivals---in practice, no more than two vessels enter the port at a given hour, and more often there is only one. A control hour is an hour with no cruise vessel arriving.

At the daily level, we create a hypothetical randomized experiment easily understandable from a policy point of view. We define $N_{t}$ as the number of cruise vessels entering Marseille port on day \textit{t}. Our treatment indicator is $W_{t}$ and takes two values:

\begin{equation}\label{eq:2}
W_{t} =
    \begin{cases}
      1 & \text{if N$_{t}=1$}\\
      0 & \text{if N$_{t}$=0}
    \end{cases}       
\end{equation}

Daily units with W$_{t}$ equal to one are considered as ``treated" while units with W$_{t}$ equal to zero belong to the control group. A treated day is a day with one cruise vessel arriving at the port. A control day is a day without any cruise vessel arriving. On average, there is around one cruise vessel entering the port each day in the initial sample. Therefore, the results of this hypothetical experiment can be interpreted as reflecting the contribution of cruise vessel traffic on an average day of the year.

Both experiments should be seen as independent from each other, as they aim at testing the pollution impact of cruise traffic at two different time frames. Whether the impact should be stronger for specific air pollutants at the hourly or daily level is ambiguous: a measurement campaign conducted near Marseille port detected an increase in hourly concentrations of pollution consistent with maritime traffic -- although they do not use vessel traffic data -- but failed to detect an effect at the daily level \citep{atmosud_quelle_2019}. It could still be argued that cruise traffic may also impact daily concentrations due to secondary pollutants taking time to form. 

In our setting, each hourly and daily unit has two continuous potential outcomes whose values range in the set of plausible pollutant concentrations in \si{\ugpcm}, Y$_{t}$(0) if W$_{t}$=0 and Y$_{t}$(1) if W$_{t}$=1. As explained in the following section, our matching algorithm approximates a pairwise randomized experiment by finding similar pairs of short-time series. To properly identify a causal effect with time series data, several assumptions are required. First, we should check that pairs are well-balanced in terms of interventions occurring in the pre-treatment period. Second, we should make the Stable Unit Treatment Value Assumption (STUVA) plausible \citep{rubin1974estimating, imbens2015matching, baccini2017assessing, forastiere2020assessing}. In the context of our hypothetical experiments, there must be no spillovers effects within and across matched pairs. Within a matched pair, a treated unit should be temporally far away from a control unit. Across pairs, the first lead outcome of a treated unit in one pair should not be used a control in another pair. This assumption could be harder to make for the hourly experiment since we do not have clear priors on when the treatment would actually occur. For instance, during the maneuvering phase, cruise vessels could already impact air pollutant concentrations before being docked. Once they are docked, they keep their engines on and could emit air pollution in the following hours. In the time series of the treated unit, it is therefore difficult to precisely define which lags and leads of the concentration of an air pollutant is not affected by vessel emission.

\subsection{Stage 2: Designing the Hypothetical Randomized Experiments}

At the design stage, our goal is to obtain a sample of similar units for which the assignment to the treatment and control groups can be assumed to be unconfounded \citep{rubin1991practical}. Formally, this unconfoundedness assumption states that the assignment to treatment is independent from the potential outcomes given a set of \textit{observed} confounders. Instead of adjusting for confounding variables with a multivariate regression model, we use a novel pair-matching algorithm to obtain treated and control units with similar values for observed covariates \citep{sommer2021assessing}. Matching is a nonparametric method which prunes the observations to limit the imbalance between treated and control units \citep{ho2007matching, rubin2006matched, stuart2010matching, imbens2015matching}. By revealing the common support available in the data, matching avoids the statistical model to extrapolate to units without empirical counterfactual. 

Concretely, let \textbf{X}$_{t}$ be the vector of observed covariates for each unit, with \textit{t} the time indicator and X$^{(k)}_{t}$ the k$^{th}$ covariate. Our algorithm matches a treated unit to a control unit only if the component-wise distances between their covariate vectors (X$^{(1)}_{t}$, X$^{(2)}_{t}$, \ldots, X$^{(K)}_{t}$) are lower than pre-defined thresholds ($\delta_{1}$, $\delta_{2}$, \ldots, $\delta_{K}$). For a pair of covariate vectors \textbf{X}$_{t}$ and \textbf{X}$_{t'}$, we use the following distance:

\begin{equation}
\Delta_{\textbf{X}_{t}, \textbf{X}_{t'}} =
    \begin{cases}
      0 & \text{if $\lvert X^{(k)}_{t} - X^{(k)}_{t'} \lvert < \delta_k$ for all k}\\
      + \infty  & \text{otherwise}
    \end{cases}       
\end{equation}

Compared to a propensity score approach, we can make sure with this algorithm that observed confounders and their lags are balanced within pairs \citep{greifer2021matching}. To limit confounding, we select two sets of covariates. First, calendar variables (i.e., hour of the day, day of the week, bank day, holidays, month, and year) are related to both vessel traffic and air pollution. Second, weather covariates (i.e., average temperature, rainfall indicator, average humidity, wind direction blowing either from the East or West, and wind speed) could also influence both vessel traffic and air pollution. We use lags of these variables to ensure that treated and control units are as similar as possible before the treatment occurs. We define matching thresholds noting that they should be strict enough to make treated and control units comparable with each other, but not too strict to avoid reducing the sample size too much. Given this trade-off, the thresholds are stricter for the hourly experiment for which the sample size is 24 times larger.  \Cref{tab_thresholds} displays all threshold values used in our matching procedure.

At the hourly level, we match exactly on calendar variables (hour of the day, day of the week, bank days, holidays) over the current and two previous hours before the treatment occurred (i.e., 0, 1, 2 lags) and allow a maximum distance of 30 days between treated and control units. For weather parameters, we carried out an iterative process, for which we tried different discrepancy values and kept the ones that led to balanced treated and control groups while resulting in enough matched pairs. We found that a maximum discrepancy of around half a standard deviation often yields a good balance. We match exactly for the East and West wind directions because they play an important role in the dispersion of air pollutants.

At the daily level, we create similar pairs of treated and control units over the current and previous day before the treatment occurred (i.e., 0 and 1 lags). We relax some of the constraints from the hourly level to have enough matched pairs. We strictly match on the day of the week, bank days, and holidays over the two days of the series. We allow treated and control units to have up to three years of difference, but they should belong to the same month. For weather parameters, we match exactly on the rainfall indicator and the wind direction on days \textit{t} and \textit{t-1}, and we allow a small discrepancy threshold for temperature and wind speed on \textit{t} and \textit{t-1}.

Based on these thresholds, each treated unit is matched to its closest control unit using a maximum bipartite matching algorithm \citep{micali1980v}. If no control unit is available to match a treated unit, it is discarded. We thus approximate the design of a pairwise randomized experiment where the assignment mechanism is a Bernoulli trial with a treatment probability of 0.5. Given this design, for each hypothetical experiment, the number of possible permutations is 2$^{P}$, with $P$ being the number of matched pairs.

\begin{table}[ht!]
\footnotesize
\caption{Maximum Discrepancies allowed for each Covariate between Treated and Control Units, Hourly and Daily Experiments.}\label{tab_thresholds}
\begin{threeparttable}
\begin{tabularx}{\textwidth}{Xcc}
\toprule
& \textbf{\normalsize Hourly Experiment} & \textbf{\normalsize Daily Experiment} \\ \midrule
\multicolumn{3}{@{}l}{\textbf{\normalsize Calendar Indicators}}\\
\rowcolor{Gray}
Distance in days                               & 30              & 1095               \\
Hour of the day in \textit{t}   & 0               &               \\
Weekday, Bank Days and Holidays in \textit{t}   & 0               & 0              \\
\rowcolor{Gray}
Weekday, Bank Days and Holidays in \textit{t-1} & 0               & 0              \\
Weekday, Bank Days and Holidays in \textit{t-2} & 0               &                \\
\rowcolor{Gray}
Month in \textit{t}                                      &           & 0 \\
\rowcolor{Gray}
\midrule
\multicolumn{3}{@{}l}{\textbf{\normalsize Weather Parameters}}\\
Average Temperature (°C) in \textit{t}                       & 4               & 4              \\
\rowcolor{Gray}
Average Temperature (°C) in \textit{t-1}                     & 4               & 4              \\
Average Temperature (°C) in \textit{t-2}                     & 4               &                \\
\rowcolor{Gray}

Rainfall Dummy in \textit{t}                            & 0               & 0              \\
Rainfall Dummy in \textit{t-1}                         & 0               & 0              \\
\rowcolor{Gray}

Rainfall Dummy in \textit{t-2}                          & 0               &                \\
Average Humidity (\%) in \textit{t}                          & 9               &              \\
\rowcolor{Gray}

Average Humidy (\%) in \textit{t-1}                         & 9               &              \\
Average Humidity (\%) in \textit{t-2}                        & 9               &                \\
\rowcolor{Gray}

Wind direction in 2 categories (East/West) \textit{t}                  & 0            &  0        \\
Wind direction in 2 categories (East/West) \textit{t-1}                  & 0            & 0           \\
\rowcolor{Gray}
Wind direction in 2 categories (East/West) \textit{t-2}                  & 0            &               \\ 
Wind speed (m/s) in \textit{t}                                & 1.8  & 2 \\
\rowcolor{Gray}
Wind speed (m/s) in \textit{t-1}                               & 1.8 & 2 \\
Wind speed (m/s) in \textit{t-2}                               & 1.8 &  \\
\bottomrule
\end{tabularx}
\begin{tablenotes}
\footnotesize
\item \textit{Notes}: This table displays the maximum distance allowed for each covariate in the pair matching algorithm, for each experiment. For example, it means that, for each matched pair, treated and control units must have the same values for weekday, bank days and holidays indicators in \textit{t}. If a discrepancy value is missing in one of the two column, it means that the associated covariate was not used for matching for the corresponding experiment.
\end{tablenotes}
\end{threeparttable}
\end{table}

\pagebreak

\subsection{Stage 3: Analyzing the Experiments using Randomization-based Inference}

Once we obtain a balance sample of matched pairs, we implement a randomization-based inference procedure to analyze the effects of cruise vessels on air pollutant concentrations. Given that we have a low number of matched pairs, we rely on this particular mode of inference since it avoids large-sample approximation and is distribution-free. The hypothetical random allocation of the treatment in the matched samples is the only source of uncertainty for inference.

\paragraph{Point estimate for the unit-level treatment effect size.}

We assume a constant additive unit-level treatment effect $\tau$:
\begin{equation}
Y_{t}(1) =  Y_{t}(0) + \tau \text{      }  \forall t =1,\ldots, \textit{T}  
\end{equation}

Under such assumption, the average pair difference in pollutant concentrations across treated and control units is an unbiased estimator for $\tau$ \citep{keele2012strengthening}. Thus, for an experiment with $I_{\mathrm{Pairs}}$ matched pairs, where $Y_{1,i}^{\mathrm{obs}}$ is the observed pollutant concentration for the treated unit of pair $i$ and $Y_{0,i}^{\mathrm{obs}}$ is the observed pollutant concentration for the control unit of pair $i$, we take as a point estimate the observed value of the average pair differences:

\begin{equation}
  \hat{\tau}=\frac{1}{I_{\mathrm{Pairs}}}\sum_{i=1}^{N_{\mathrm{Pairs}}} (Y_{1,i}^{\mathrm{obs}}-Y_{0,i}^{\mathrm{obs}})  
\end{equation}

\paragraph{Randomization-based quantification of uncertainty.}

We carry out a test-inversion procedure to build 95\% Fisherian (also called ``Fiducial") Intervals (FI) for the constant unit-level treatment effect. We closely follow the procedure detailed by T. Dasgupta and D.B. Rubin in their forthcoming book \citep{dasgupta_rubin_2021}. On our \href{
https://lzabrocki.github.io/cruise_air_pollution/3_toy_example_randomization_inference.html}{website}, we provide a detailed toy example to explain this mode of inference. Instead of gauging a null effect for all units, we test \textit{J} sharp null hypotheses $H_{0}^{j}$: Y$_{t}$(1) =  Y$_{t}$(0) + $\tau_{j}$ for j =1,$\ldots$, \textit{J}, where $\tau_{j}$ represents a constant unit-level treatment effect size. We test a sequence of sharp null hypotheses of constant treatment effects ranging from -10 \si{\ugpcm} to +10 \si{\ugpcm} with an increment of 0.1 \si{\ugpcm}. As a test-statistic, we use the sample average of pair differences, which is commonly used in randomization-based inference \citep{keele2012strengthening, imbens2015causal}. For each constant treatment effect \textit{j}, we calculate the upper \textit{p}-value associated with the hypothesis $H_{0}^{j}$: Y$_{t}$(1) - Y$_{t}$(0) $>$ $\tau_{j}$ and the lower \textit{p}-value for $H_{0}^{j}$: Y$_{t}$(1) - Y$_{t}$(0) $<$ $\tau_{j}$. We run 10,000 permutations for each hypothesis to approximate the null distribution of the test statistic. Running the exact number of possible allocations is computationally too intensive given the number of matched pairs we found. The results of testing the sequence of \textit{J} hypotheses $H_{0}^{j}$: Y$_{t}$(1) - Y$_{t}$(0) $>$ $\tau_{j}$ forms an upper \textit{p}-value function of $\tau$, $p^{+}(\tau)$, while the sequence of alternative hypotheses $H_{0}^{j}$: Y$_{t}$(1) - Y$_{t}$(0) $<$ $\tau_{j}$ makes a lower \textit{p}-value function of $\tau$, $p^{-}(\tau)$. To calculate the bounds of the 100(1-$\alpha$)\% Fisherian interval, we solve $p^{+}(\tau) = \frac{\alpha}{2}$ for $\tau$ to get the lower limit and $p^{-}(\tau) = \frac{\alpha}{2}$ for the upper limit. We set our $\alpha$ significance level to 0.05, and thus calculate two-sided 95\% Fisherian intervals. This procedure allows us to get the range of constant treatment effects consistent with our data and with the hypothetical assignment mechanism we posit \citep{rosenbaum2010design, dasgupta2015}. 

One limit of randomization inference is that it assumes a constant treatment effect across units. In our study, this is probably an unrealistic assumption: for example, we expect weather conditions to affect the relationship between cruise traffic and pollution concentration. To overcome this limit, we implement two other inference procedures that are designed for heterogeneous treatment effects. First, we can compare the results of the randomization inference procedure with the ones we would obtain with Neyman's approach \citep{neyman1923applications}. In that case, the inference procedure is built to target the average causal effect and the source of inference is both the randomization of the treatment and the sampling from a population. We can estimate the finite sample average effect, $\tau_{\text{fs}}$, with the average of observed pair differences $\hat{\tau}$, defined as:

\begin{equation*}
  \hat{\tau} = \frac{1}{I}\sum_{i=1}^I(Y^{\text{obs}}_{\text{t},i}-Y^{\text{obs}}_{\text{c},i}) = \overline{Y}^{\text{obs}}_{\text{t}} - \overline{Y}^{\text{obs}}_{\text{c}}
\end{equation*}

\noindent Here, the subscripts $t$ and $c$ respectively indicate if the unit in a given pair is treated or control. $I$ is the number of pairs. Since there are only one treated and one control unit within each pair, the standard estimate for the sampling variance of the average of pair differences is not defined. We can however compute a conservative estimate of the variance, as explained in chapter 10 of \cite{imbens2015causal}:

\begin{equation*}
  \hat{\mathbb{V}}(\hat{\tau}) = \frac{1}{I(I-1)}\sum_{I=1}^I(Y^{\text{obs}}_{\text{t},i}-Y^{\text{obs}}_{\text{c},i} - \hat{\tau})^{2}
\end{equation*}

\noindent We finally compute an asymptotic 95\% confidence interval using a Gaussian distribution approximation:

\begin{equation*}
\text{CI}_{0.95}(\tau_{\text{fs}}) =\Big( \hat{\tau} - 1.96\times \sqrt{\hat{\mathbb{V}}(\hat{\tau})},\; \hat{\tau} + 1.96\times \sqrt{\hat{\mathbb{V}}(\hat{\tau})}\Big)
\end{equation*}

Second, we implement the randomized inference approach recently developed by \cite{wu2021randomization} that is asymptotically conservative heterogeneous effects. The procedure follows exactly the same steps previously described but is based on a studentized test statistic that is equal to the observed average of pair differences divided by Neyman's standard error of a pairwise experiment. 




\section{Results}\label{sec:results}

In this section, we first present covariate balance diagnostics on the performance of our matching procedure. We then display and interpret the results for the effects of hourly and daily cruise vessel traffic on air pollutant concentrations. We end the section with a set of robustness checks.


\subsection{Matching Results}

\begin{table}[!ht]
\caption{Number of Matched Pairs by Experiment.}\label{matching_results}
\begin{threeparttable}
\begin{tabularx}{\textwidth}{Xcc}
\toprule
& \textbf{\shortstack{Hourly \\ Cruise Experiment}} 
& \textbf{\shortstack{Daily \\ Cruise Experiment}}  \\ \midrule
\rowcolor{Gray}
N$_{\mathrm{Total}}$ &  96,432  & 4,018  \\
N$_{\mathrm{Treated}}$ &  4,034   & 2,485  \\
\rowcolor{Gray}
N$_{\mathrm{Control}}$ & 92,396   & 1,532  \\
N$_{\mathrm{Pairs}}$   & 138   & 189   \\
\bottomrule
\end{tabularx}
\begin{tablenotes}
\footnotesize
\item \textit{Notes}: This table displays the total number of observations, N$_{\mathrm{Total}}$ for each experiment, the number of potential treated and controls units before matching, N$_{\mathrm{Treated}}$ and N$_{\mathrm{Control}}$, and the number of matched pairs, N$_{\mathrm{Pairs}}$.
\end{tablenotes}
\end{threeparttable}
\end{table}

\paragraph{Hourly matching diagnostics.} As shown in \Cref{matching_results}, our matching procedure at the hourly level results in few matched treated units, with less than 4\% of treated units matched to similar control units. Two main reasons explain this result. First, cruise vessel traffic is regular over time, so that it is hard to find similar control and treated hours which are not temporally too far away from each other. Second, even if we relax our matching constraints, it is difficult to find treated and control units with similar weather covariates. We check that within pairs spillovers are not likely to occur since within a pair, treated and control units are at least 7 days away. However, there could be spillovers across pairs. For instance, for 16\% of treated units, the minimum distance with a control unit in another pair is inferior or equal to 5 hours. Dropping these pairs or modifying the matching algorithm to avoid having pairs too close temporally of each others would be required to avoid spillover effects.

\begin{figure}[!ht]
\caption{Intervention Diagnostics for the Hourly Experiment.}
    \centering\includegraphics[scale=0.5]{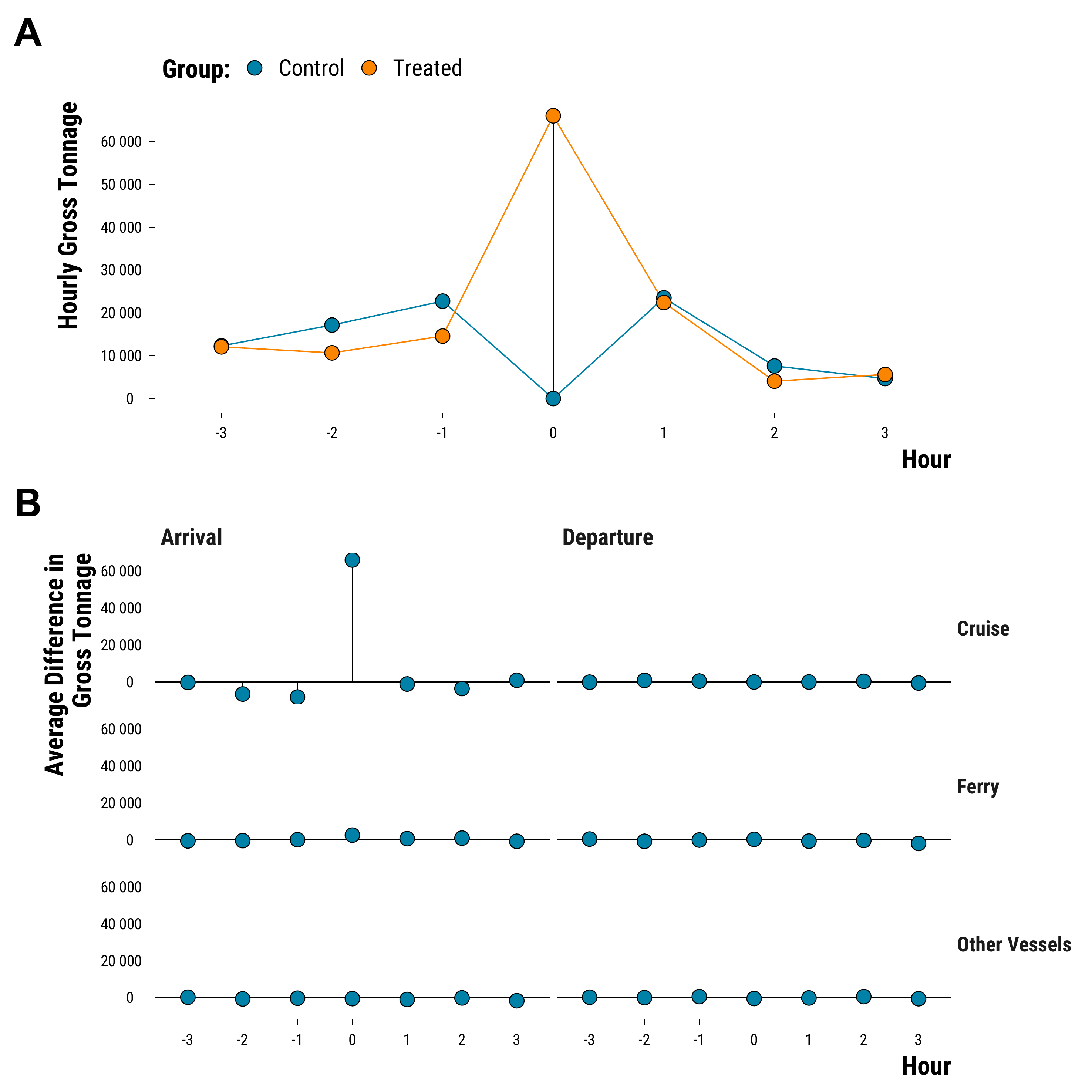}
\label{fig:diagnostics_matching_intervention_hourly}
\caption*{\textit{Notes:} Panel A shows the average hourly total gross tonnage for matched treated and control units. Panel B plots the average difference in total gross tonnage between treated and control units by vessel type and flow.}
\end{figure}

\begin{figure}[!ht]
\caption{Improvement in Covariates Balance for the Hourly Experiment.}
    \centering\includegraphics[scale=0.8]{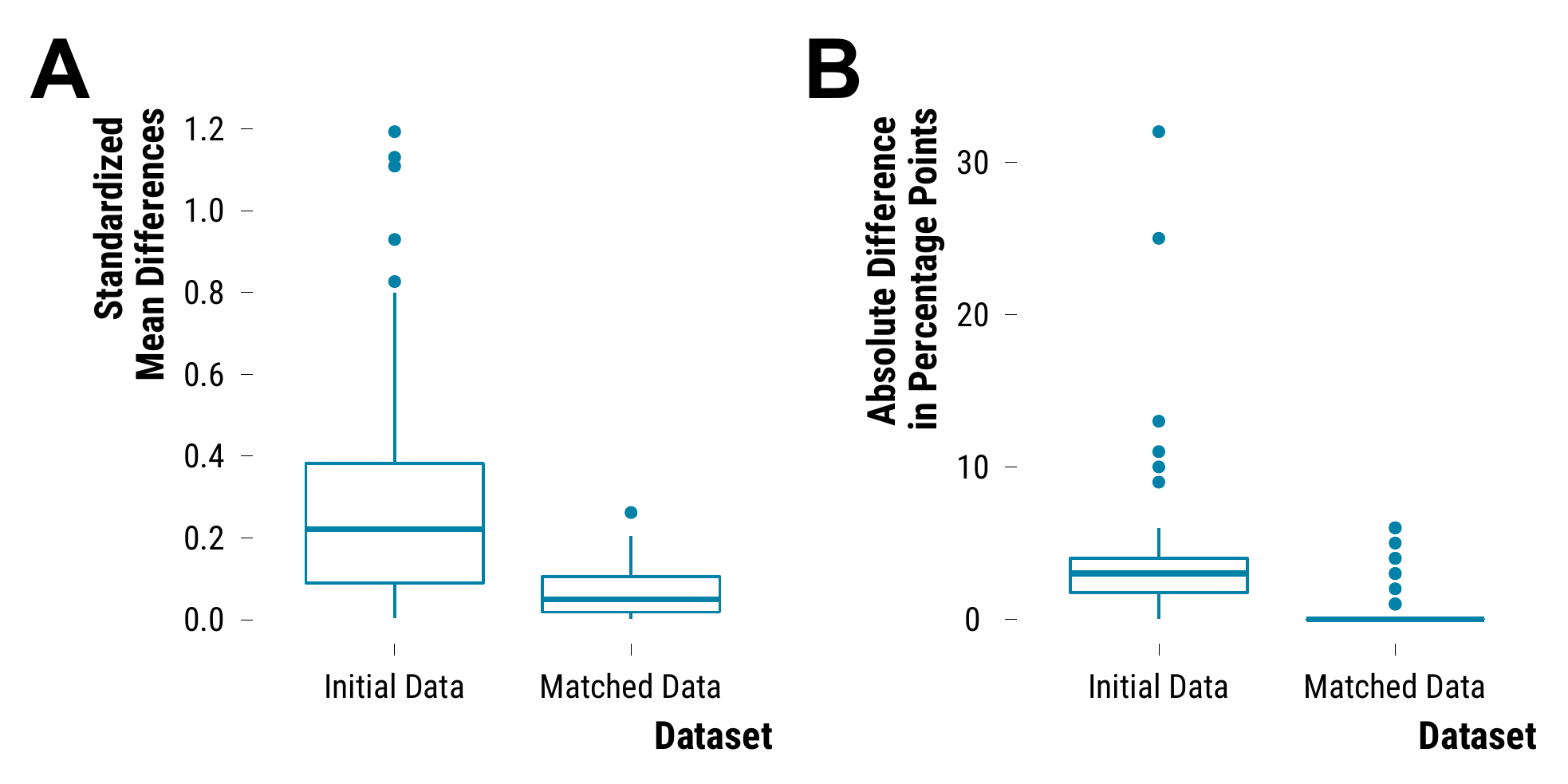}
\label{fig:diagnostics_matching_balance_hourly}
\caption*{\textit{Notes:} Panel A shows the boxplot distribution of the absolute standardized mean differences in continuous covariates before and after matching. Panel B shows the boxplot distribution of the absolute mean differences in categorical covariates before and after matching.}
\end{figure}

In \Cref{fig:diagnostics_matching_intervention_hourly}, Panel A displays the average increase in cruise vessel arrivals at hour 0. The average difference in gross tonnage between treated and control units is about 65,000 for the hourly cruise experiment, which is the average gross tonnage of one cruise vessel. Panel B shows that, on average, treated and control units have similar vessel traffic for other vessel types and flows. The matching procedure at the hourly level improves the overall balance of covariates as shown in \Cref{fig:diagnostics_matching_balance_hourly}. Further diagnostics on covariates balance are available at the hourly level on our \href{https://lzabrocki.github.io/cruise_air_pollution/1_7_hourly_checking_balance_improvement.html}{website}.

Finally, we compare the matched data to the initial data. Matched hours are more likely to fall on spring and summer days, which are hotter on average. They fall disproportionately around 7 am, the time where cruise vessels tend to arrive in Marseille port. All comparisons are available on our \href{https://lzabrocki.github.io/cruise_air_pollution/1_4_hourly_comparing_matching_matched_data.html}{website}.

\paragraph{Daily matching diagnostics.} At the daily level, we found 189 matched pairs, which means that 8\% of the treated units were matched to similar control units. There should not be within pair spillovers since treated and control units are at least 7 days away. However, as in the hourly experiment, there could be spillovers across pairs if cruise vessel emissions impact the first lead of air pollutant concentrations. For 22\% of treated units, the minimum distance with a control unit in another pair is equal to one day. 

\begin{figure}[!ht]
\caption{Intervention Diagnostics for the Daily Experiment.}
    \centering\includegraphics[scale=0.5]{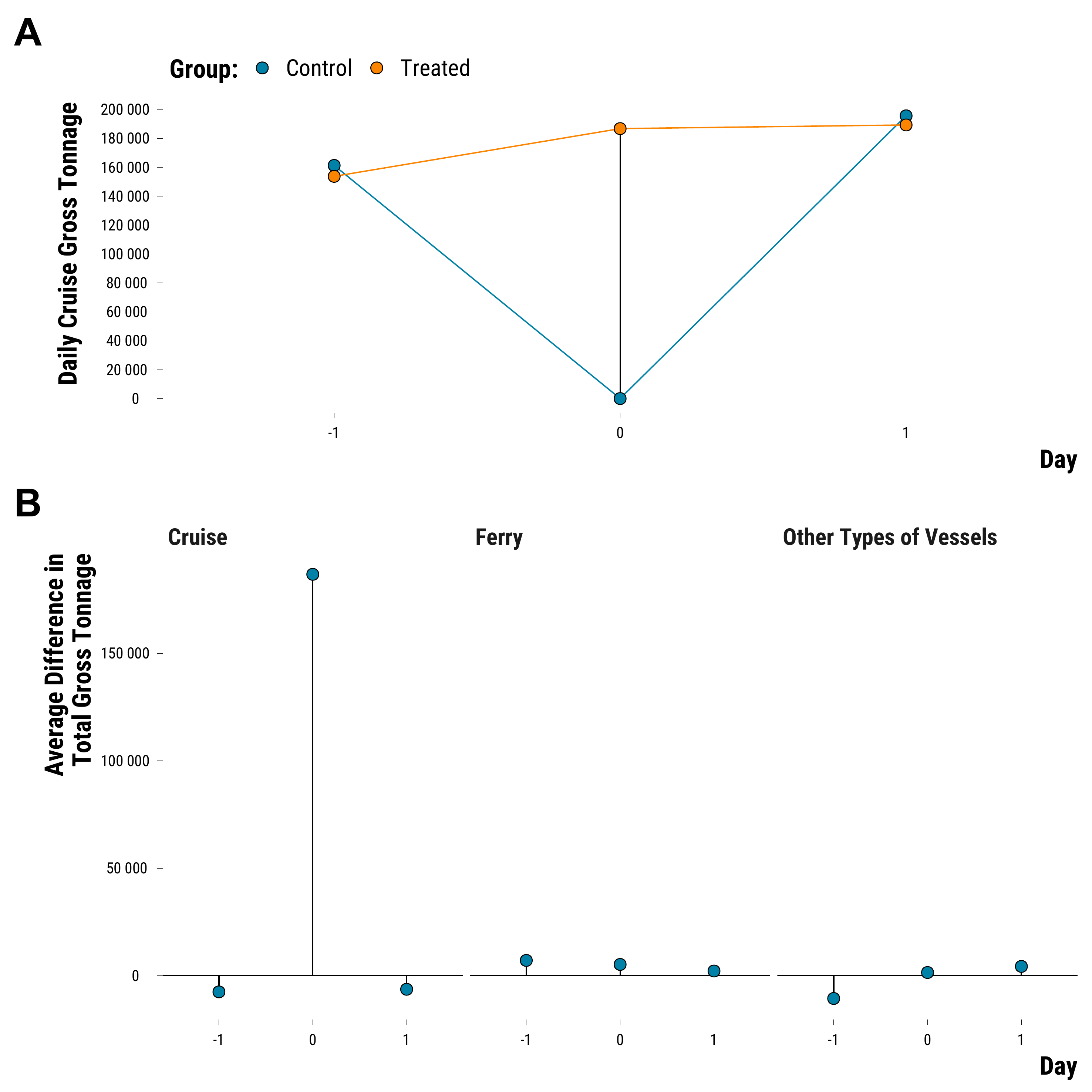}
\label{fig:diagnostics_matching_intervention_daily}
\caption*{\textit{Notes:} Panel A shows the average daily total gross tonnage for matched treated and control units. Panel B plots the average difference in total gross tonnage between treated and control units by vessel type and flow.}
\end{figure}

\begin{figure}[!ht]
    \caption{Improvement in Covariates Balance for the Daily Experiment.}
    \centering\includegraphics[scale=0.8]{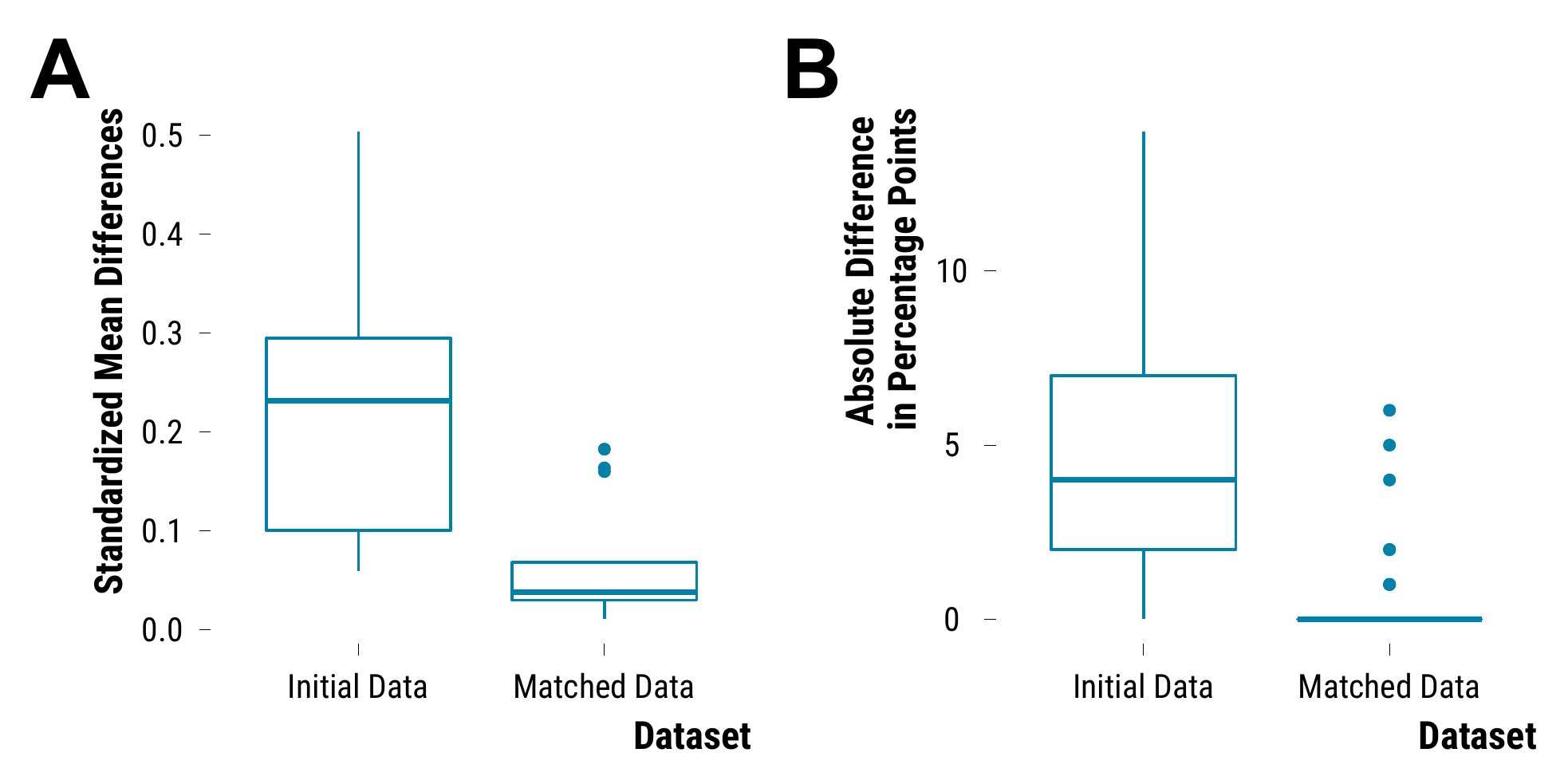}
\label{fig:diagnostics_matching_balance_daily}
\caption*{\textit{Notes:} Panel A shows the boxplot distribution of the absolute standardized mean differences in continuous covariates before and after matching. Panel B shows the boxplot distribution of the absolute mean differences in categorical covariates before and after matching.}
\end{figure}

In Panel A of \Cref{fig:diagnostics_matching_intervention_daily}, the average difference in gross tonnage between treated and control units is around 150,000, which corresponds to the tonnage of two vessels. The cruise vessel entering the port in the morning most likely leaves the port in the evening after docking at the port during the day. The variation in gross tonnage for other vessel types is similar across treated and control units, as shown in Panel B of \Cref{fig:diagnostics_matching_intervention_daily}. Similarly to the hourly experiment, the matching procedure improved the balance of covariates, as shown in \Cref{fig:diagnostics_matching_balance_daily}. Further diagnostics on covariates balance are available at the daily level on our \href{https://lzabrocki.github.io/cruise_air_pollution/2_7_daily_checking_balance_improvement.html}{website}.

As for the hourly experiment, days in the matched sample are more often in summer so that they are hotter, less rainy and with a lower wind speed than the average day from the initial sample. All comparisons are available on our \href{https://lzabrocki.github.io/cruise_air_pollution/2_4_daily_comparing_matching_matched_data.html}{website}.


\subsection{The Effects of Cruise Vessel Traffic on Air Pollutants}

\begin{figure}[!ht]
\centering
\caption{Effects of Cruise Vessel Traffic on Pollutant Concentrations at the Hourly Level.}
\includegraphics[width=\linewidth]{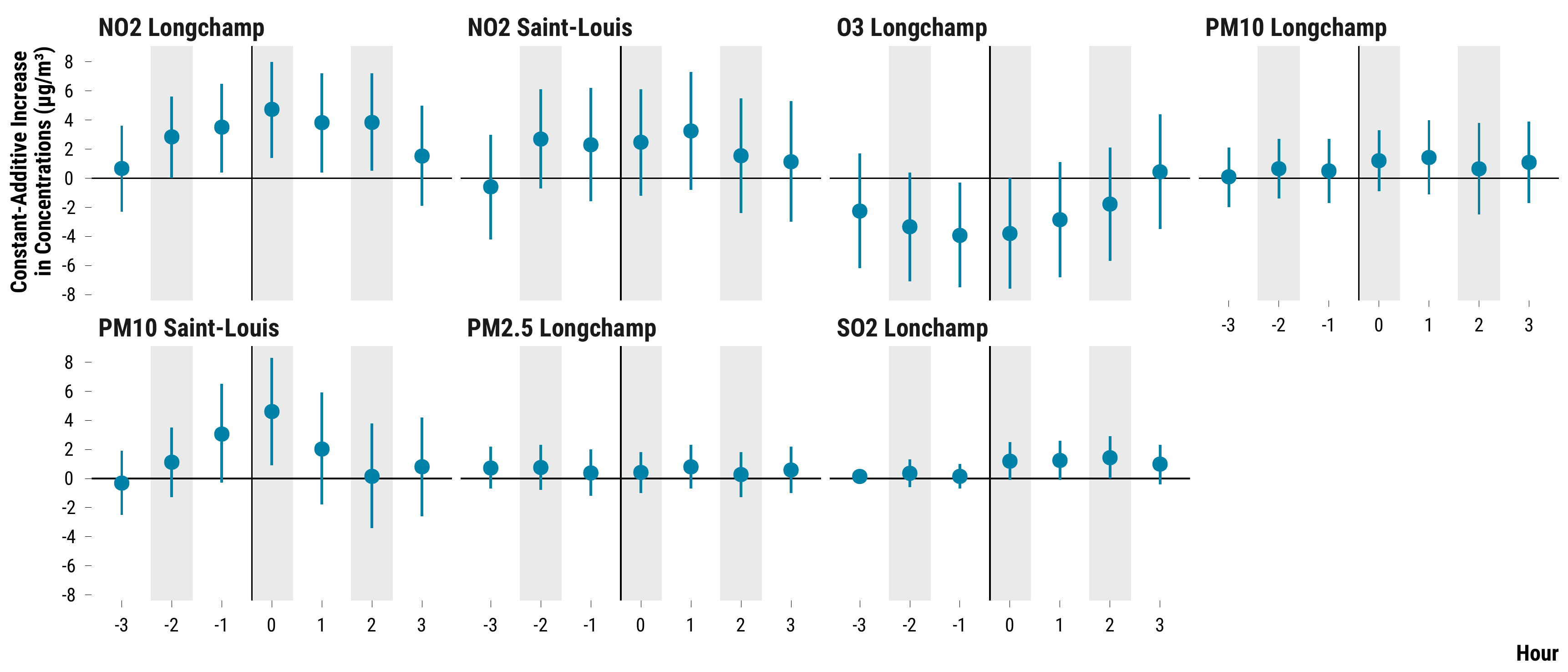}
\label{fig:graph_ci_pollution_hourly}
\caption*{\textit{Notes:} The treatment occurs at hour 0. Dots represent the point estimate of the unit-level treatment effect on a pollutant concentration. Lines are 95\% Fisherian intervals of constant treatment effects consistent with the data. The effects are plotted from the third lag to the third lead.}
\end{figure}

\paragraph{Hourly Effects.} In \Cref{fig:graph_ci_pollution_hourly}, we plot the point estimates and the 95\% Fisherian intervals of the constant treatment effects on air pollutant concentrations that are consistent with our data. We compute these effects for the three previous hours before the treatment occurs up to the three following hours in order to capture the impacts of emissions during the maneuvering phase of cruise vessels but also while they are docked with their engines on.

For NO$_{2}$, we observe an increase in concentration from the second previous hour up to the second following hour. The pattern is clearer for the Longchamp station than the Saint-Louis station where the signal is more noisy. At hour 0, concentrations are higher by 4.7 \si{\ugpcm} (95\% FI: [1.4, 8.0]). In relative terms, this represents a 16\% increase in the average hourly concentration of NO$_{2}$ measured at Longchamp station. The 95\% Fisherian are relatively wide since the data are consistent with constant effects ranging from a 5\% increase up to a 27\% increase in concentration.

For O$_{3}$, we see the opposite relationship since there seems to be a decrease in concentration in the three previous hours, followed by an increase in the following hours. In hour 0, there is a constant decrease of O$_{3}$ concentrations by 3.8 \si{\ugpcm} (95\% FI: [-7.6, 0.0]). This is equivalent to a 7\% decrease in the average hourly concentration of the air pollutant. Again, the 95\% Fisherian intervals are wide: the data are consistent with null effects up to a 14\% decrease.

For SO$_{2}$, we observe an increase in concentrations of 1.2 \si{\ugpcm} at hour 0 (95\% FI: [-0.1, 2.5]), which persists over the two following hours. The constant increase is equivalent to a very large relative increase in concentration by 52\%. The 95\% Fisherian interval is also wide since the data are consistent with a relative decrease of 4\% up to a relative increase of 109\%.

For particulate matter,  we observe an increase in PM$_{10}$ concentrations measured at Saint-Louis that is followed by a decrease. At hour 0, the constant increase is equal to 4.6 \si{\ugpcm} (95\% FI: [0.9, 8.3]). This is equivalent to a 15\%: the data are consistent with relative increase from 3\% up to 27\%. There are no very clear patterns for PM$_{10}$ and PM$_{2.5}$ concentrations measured at Longchamp station. 

\pagebreak

\begin{figure}[!ht]
\centering
\caption{Effects of Cruise Vessel Traffic on Pollutant Concentrations at the Daily Level.}
\includegraphics[width=\linewidth]{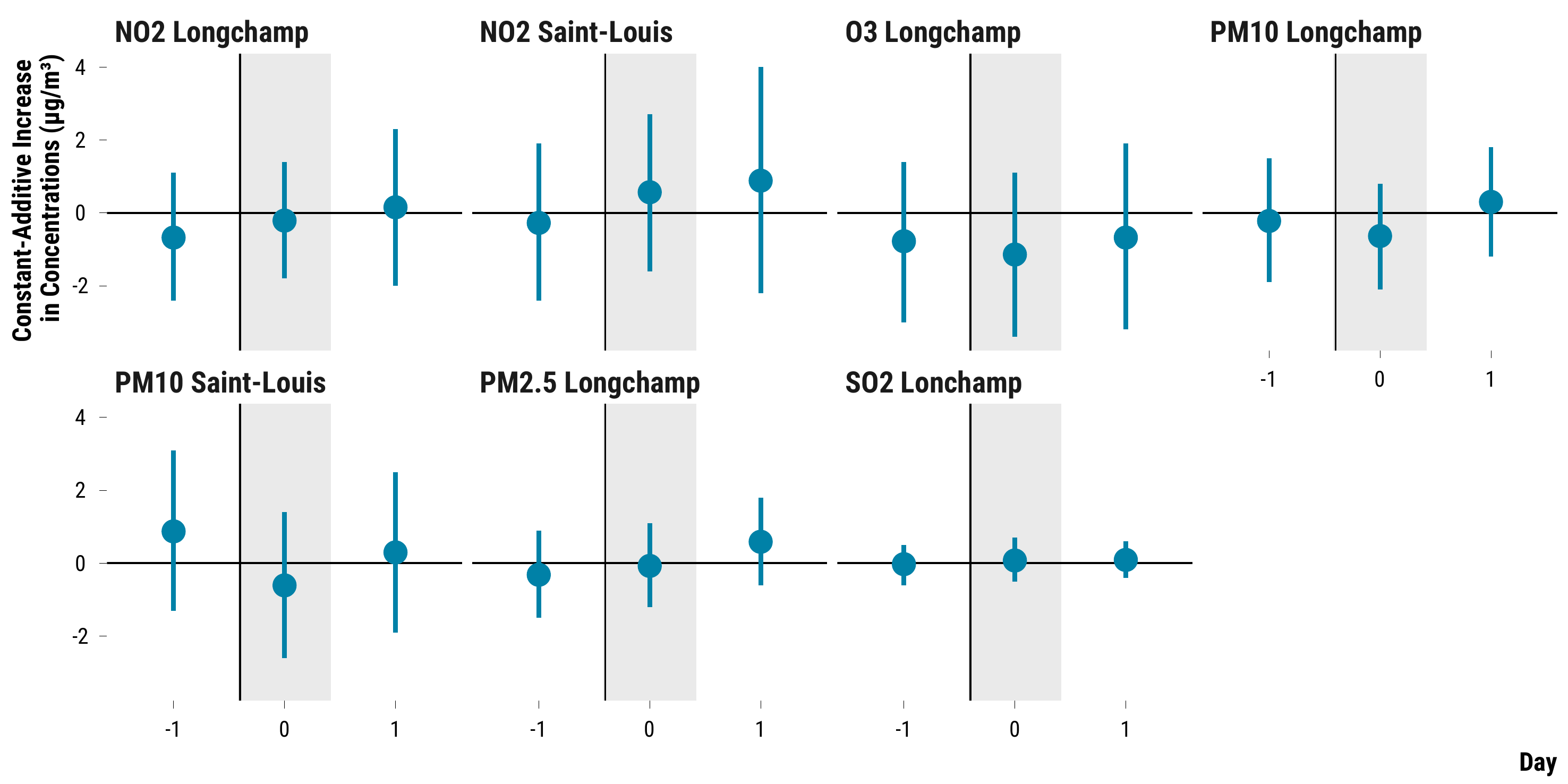}
\label{fig:graph_ci_pollution_daily}
\caption*{\textit{Notes:} The treatment occurs at day 0. Dots represent the point estimate of the unit-level treatment effect on a pollutant concentration. Lines are 95\% Fisherian intervals of constant treatment effects consistent with the data. The effects are plotted from the first lag to the first lead.}
\end{figure}

\paragraph{Daily Effects.} \Cref{fig:graph_ci_pollution_daily} shows the results for the daily experiment. We can see point estimates are nearly null for all air pollutants. The 95\% Fisherian intervals are however however relatively large. For instance, if the point estimate for the constant effect on NO$_{2}$ in Longchamp is nearly null, the data are consistent with effects ranging from a 6\% decrease up to a 5\% increase.

\paragraph{Neyman's approach and randomization inference for average treatment effects.} For the hourly and daily experiments, the 95\% Fisherian intervals for constant treatment effects are very similar to the intervals for the average treatment effects computed with Neyman's approach (see \Cref{fig:graph_ci_neyman_daily}. They are also very similar to those found with the studentized randomization inference that is conservative for weak null hypotheses (see \Cref{fig:graph_ci_fisher_weak_daily}. With these two alternative mode of inference, we can also confidently interpret the previous 95\% Fisherian intervals as the range of \textit{average} treatment effects consistent with the data. 

\paragraph{Heterogeneity Analysis.} We carry out two heterogeneity analyses for the \href{https://lzabrocki.github.io/cruise_air_pollution/1_8_hourly_analysis_results.html#heterogeneity-analysis}{hourly} and \href{https://lzabrocki.github.io/cruise_air_pollution/2_8_daily_analysis_results.html#heterogeneity-analysis}{daily} experiments. First, depending on the wind direction, the effects of cruise vessel emissions on air pollutant concentrations are likely to be attenuated or increased. At the hourly level, we observe stronger differences in concentrations for all air pollutant when the wind is blowing from the West, that is to say when vessel emissions are more likely to spread over the city (see \Cref{fig:graph_wd_hourly}). At the daily level, there are no clear patterns. Second, we also visually explore the relationship between pair differences in air pollutant concentrations against the pair differences in gross tonnage. Ideally, we should see a positive relationship since the higher the pair difference in gross tonnage (i.e., the higher the treatment shock is), the larger the pair difference in concentrations should be. We do not see any clear patterns, both for the hourly and daily experiments.


\subsection{Robustness checks}

We carry out several robustness checks to evaluate different aspects of the design and results of our study.

\paragraph{Randomization check for overall balance.} During the matching procedure, we assess the balance with Love plots that display for each covariate the standardized difference in means between treated and control units before and after matching. To better assess the overall balance, we implement the randomization inference method developed by \cite{branson2021randomization} to evaluate if the treatment indicator is as-if randomized according the pairwise structure in the matched data. As a test statistic, \cite{branson2021randomization} proposes to use the Mahalanobis distance which summarizes the imbalance in the means of all covariates but also takes into account their joint relationships. The randomization inference procedure consists in permuting the treatment indicator many times, computing the Mahalanobis distance for each iteration and plotting the resulting null distribution of the test statistic. If the observed value of the Mahalanobis distance is far away from the distribution, it means that the treatment is not as-if randomized according to observed covariates. For both the \href{https://lzabrocki.github.io/cruise_air_pollution/1_7_hourly_checking_balance_improvement.html#randomization-check-for-covariate-balance
}{hourly} and \href{https://lzabrocki.github.io/cruise_air_pollution/2_7_daily_checking_balance_improvement.html#randomization-check-for-covariate-balance
}{daily} experiments, we find clear evidence that the treatment is randomized according to a pairwise structure.

\paragraph{Sensitivity to hidden bias.} The causal interpretation of our results is based on the plausibility of the hypothetical experiment and the unconfoundedness assumption \citep{rubin1991practical}. This is a strong assumption as it states that the treatment assignment probability is not a function of potential outcomes given observed and \textit{unobserved} counfounding factors \citep{sekhon2009opiates}. To evaluate the consequence of hidden bias, we rely on the sensitivity analysis framework developed by \citep{rosenbaum1987sensitivity, rosenbaum2010design}. The goal of this method is to quantify how the treatment estimates would be altered by the effect of an unobserved confounder on the treatment odds, denoted by $\Gamma$. In our matched pairwise experiments, we make the assumption that within each pair, control and treated units have the same probability of 0.5 to be treated, that is to say to have a positive shock on cruise traffic. The odds of treatment is such that $\Gamma = (0.5/(1-0.5))/(0.5/(1-0.5)) =1$. As explained in \cite{rosenbaum2010design}, we can implement a randomization inference procedure to compute the 95\% Fisherian intervals obtained for a given value of bias that the unmeasured confounder has on the treatment assignment. For instance, if we assume that an unmeasured confounder has a small effect on the odds of treatment (i.e., for a $\Gamma$ > 1 and close to 1) but the resulting 95\% Fisherian interval is consistent with negative, null and positive effects, it would imply that our results are highly sensitive to hidden bias. Conversely, if we assume that an unmeasured confounder has a strong effect on the odds of treatment (i.e., for a large $\Gamma$) and we find that the resulting 95\% Fisherian interval remains similar, it would strength our view that our results do not suffer from hidden bias. Again, the method of \cite{rosenbaum2010design} relies on the assumption of constant additive treatment effects, which is unrealistic in our study. To overcome this limit, we implement the new method proposed by \cite{fogarty2020studentized} which extends the sensitivity analysis for sample average treatment effects. In a complementary evaluation of hidden bias, we also check whether unmeasured biases could be present by using the lags of air pollutant concentrations as placebo/control outcomes \citep{imbens2015causal}. If our matched pairs are indeed similar in terms of unobserved covariates, the treatment occurring in $t$ should not influence concentration of air pollutants in the first lag at the daily level and concentrations for further lags at the hourly level.

Our sensitivity analysis reveals that the estimated effects of cruise vessels emissions on NO$_{2}$ concentrations at Longchamp station and PM$_{10}$ concentrations at Saint-Louis station could be affected by a relatively weak hidden bias. Concretely, if we fail in our matching procedure to adjust for an unobserved confounder which is 1.5 times more common among treated units, the resulting 95\% Fisherian interval for the effects on NO$_{2}$ ranges from -1.5 \si{\ugpcm} to  11.4 \si{\ugpcm} and the intervals for the effects on PM$_{10}$ ranges from -1.9 \si{\ugpcm} to 12.2 \si{\ugpcm}. Our data would be still consistent with mostly positive effects of cruise vessel emissions on these two air pollutants but they could be null and even negative. It is however hard to think about an unobserved confounder which would change the odds of treatment by 50\%. To complement this sensitivity analysis, we also note that there are no differences in the first lag of air pollutant concentrations for the daily experiment. For the hourly experiment, we also see that for further lags and leads, estimated differences in air pollutant concentration are mostly null.

\paragraph{Sensitivity of results to outliers and missing observations.} In the matched data, the observed concentration of air pollutants are sometimes very high. To make sure that our results are not influenced by outliers, we run again our randomization inference procedure with the Wilcoxon signed-rank statistic. The 95\% Fisherian intervals obtained with this test statistic are similar to those obtained with the average of pair differences (see \href{https://lzabrocki.github.io/cruise_air_pollution/1_8_hourly_analysis_results.html#outliers}{hourly} and \href{https://lzabrocki.github.io/cruise_air_pollution/2_8_daily_analysis_results.html#outliers}{daily} results). Besides, we imputed missing values and we could fear that their imputations affect the results. For instance, at the hourly, up to 25\% of the pairs have missing values for an air pollutant. Our simulation exercise also shows that large imputation errors sometimes occur. We therefore run again our randomization inference procedure for pairs with observed air pollutant concentrations: we find similar results with slightly wider 95\% Fisherian intervals (see \href{https://lzabrocki.github.io/cruise_air_pollution/1_8_hourly_analysis_results.html#missing-outcomes}{hourly} and \href{https://lzabrocki.github.io/cruise_air_pollution/2_8_daily_analysis_results.html#missing-outcomes}{daily} results).

\paragraph{Indirect treatment effect of cruise traffic.} One issue of our design could be the presence of an indirect treatment effect due to the increase in road traffic induced by cruise vessel passengers and its subsequent effects on air pollution. This is part of the causal effect that we want to capture but it is not the proper causal effect of cruise vessel emissions. We therefore check if road traffic measures are balanced before and after the treatment occurs For the hourly and daily experiments, we observe that road traffic flow and road occupancy rate appear relatively balanced across treated and control units in the matched samples of the two experiments (see \href{https://lzabrocki.github.io/cruise_air_pollution/1_7_hourly_checking_balance_improvement.html#road-traffic}{hourly balance checks} and \href{https://lzabrocki.github.io/cruise_air_pollution/2_7_daily_checking_balance_improvement.html#road-traffic}{daily balance checks}). It is the case before and after the treatment occurs: when we observe an increase in air pollutant concentrations, this is unlikely to be due to an increase in road traffic.

\paragraph{Low statistical power and inflation of statistically significant estimates.} In the hourly and daily experiments, our matching procedure resulted in few matched pairs, which decreases the precision of our treatment effect estimates. Moreover, if our statistical power is low and we obtain a "statistically significant" effect, we have a higher chance that this estimate is of the wrong sign (Type S error) and overestimates the true effect of vessel traffic on air pollutant concentrations (Type M error) \citep{gelman2014beyond, gelman2020regression}. We therefore carry out retrospective power calculations to evaluate the risks of making type-S and type-M errors. While it is impossible to know what the true effect of cruise vessels on an air pollutant is, we can calculate the statistical power and the risks to make type S and M errors under a set of plausible effect sizes using the closed-form expression derived by \cite{lu2019note} and implemented in the \texttt{retrodesign} R package by \cite{retrodesign}.

\begin{figure}[!ht]
\centering
\caption{Statistical Power, Type M and S Errors for Hourly Experiment Effect on NO$_{2}$.}
\includegraphics[width=\linewidth]{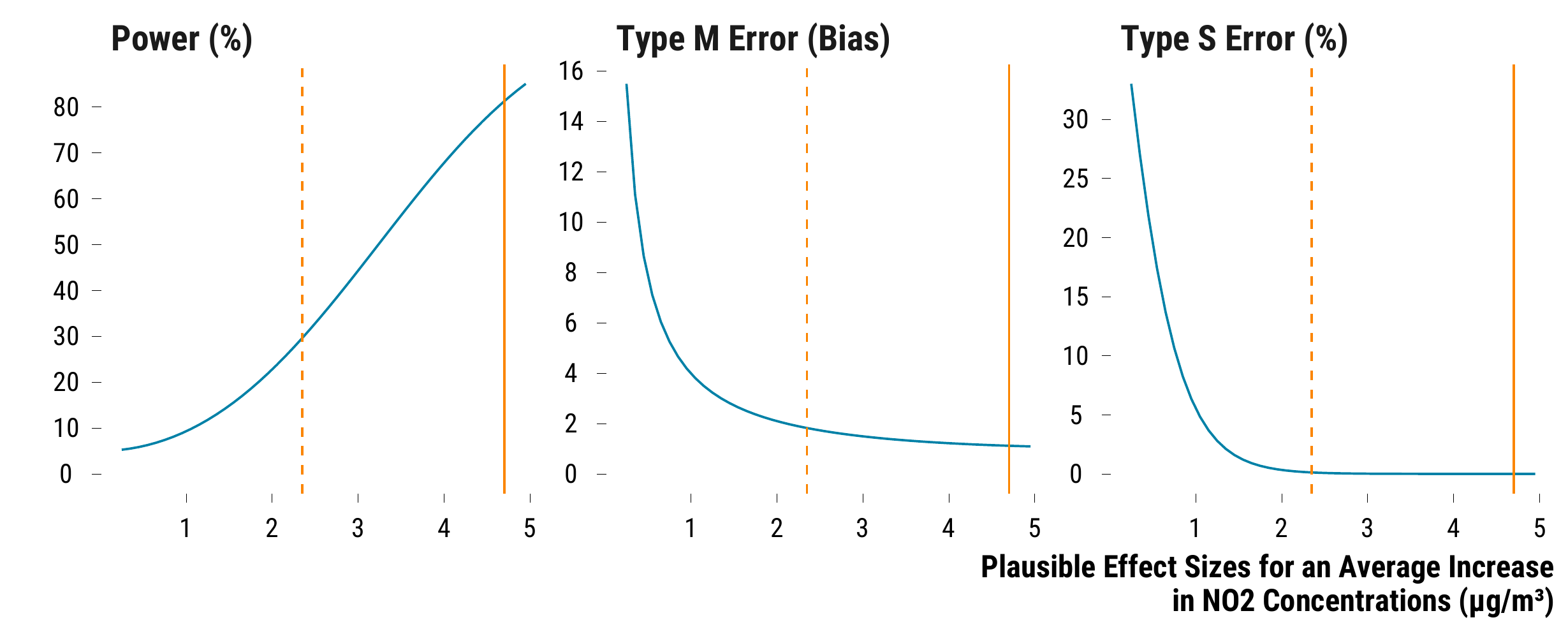}
\label{fig:graph_retrodesign}
\caption*{\textit{Notes:} In the first left panel, the statistical power of the hourly experiment on the effect of cruise vessel on NO$_{2}$ is plotted against hypothetical true effect sizes. In the middle panel, the inflation/exaggeration ratio of statistically significant estimates is plotted against against hypothetical true effect sizes. In the right panel, the probability to get a negative statistically estimate is plotted against against hypothetical true effect sizes. The solid line is the observed estimate of the average treatment effect. The dashed line is half the value of the observed estimate.}
\end{figure}

For instance, we observe a 4.7 \si{\ugpcm} increase in NO$_{2}$ concentrations in Longchamp due to cruise vessel arrivals. If other researchers think that this effect size is too large, we can retrospectively compute the power of our study according to a range of alternative true effect size. In \Cref{fig:graph_retrodesign}, if we assume that the true effect is equal to half the estimated effect size, that is to say +2.35 \si{\ugpcm} (dashed line), our study would have a power of 30\% and "statistically significant" estimates would be on average 1.8 times too large. However, the probability that a "statistically significant" estimate is of the opposite sign is nearly null. For other air pollutants for which 95\% Fisherian intervals are wider, this risk could be high. With the few number of matched pairs found in our hypothetical experiments, there is a chance that "statistically significant" estimates could be misleading: as we did, we should rather interpret the width of the 95\% Fisherian intervals.

\pagebreak

\paragraph{Strictness of the matching procedure.}  Our matching procedure is strict and result in a small number of matched pairs, both at the hourly and daily levels. This is partly due to the regularity in vessel traffic which makes it hard to find control units that are temporally close to treated units and with similar covariate values. To relax the stringency of our matching procedure, we implement a propensity score matching procedure where each treated unit is matched to its closest control unit if their distance is less than 0.01 of the standard deviation of the propensity score distribution. For instance, at the \href{https://lzabrocki.github.io/cruise_air_pollution/2_8_daily_analysis_results.html#alternative-matching-procedure}{daily} level, 1,846 pairs were matched. The Love plot indicates that covariates balance has increased but the randomization balance check suggests that the treatment is not as-if randomized according to observed covariates in the matched data. While estimates are more precise, they are relatively consistent with those found with our approach. It is very important to remind that when we compare the results of our constrained pair matching procedure with the propensity score approach, we are comparing two different subsamples of the initial dataset.

\paragraph{Comparison with an outcome regression approach.} Finally, we compare our results to estimates found using a simple multivariate regression model on the initial hourly and daily datasets. The matched datsets are selected sub-samples of the initial datasets and have different covariate values. Estimated effects are therefore not directly comparable. For each experiment, we run the following model:

\begin{equation*}
p_{t+j}=\alpha+\beta W_{t}+\textbf{X}_{t}\gamma+\textbf{C}_{t}\theta+\epsilon_{t}
\end{equation*}

\noindent where $j$ is the index of the lag or lead, \textit{t} is either the hour (for the two hourly experiment) or the day index (for the daily experiment), $p_{t+j}$ the concentration of an air pollutant $p$ at date $t+j$, $W_{t}$ the binary treatment indicator, $\textbf{X}_{t}$ the vector of weather covariates, which include the average temperature, the squared of the average temperature, an indicator for the occurrence of rainfall, the average humidity, the wind speed, the wind direction divided in the four principal directions (North-East, South-East, South-West, North-West), $\textbf{C}_{t}$ the vector of calendar variables, which are indicators for the hour of the day (for the hourly experiment), the day of the week, bank days, holidays, month, year and the interaction of these last two variables, and $\epsilon_{t}$ an error term. We run this simple model from lag 3 to lead 3 of an air pollutant for the hourly experiment on vessels' arrivals and from lag 1 to lead 1 for the daily experiment. 

\begin{figure}[!ht]
\centering
\caption{Comparison of Matching and Regression Results for the Hourly Experiment.}
\includegraphics[width=\linewidth]{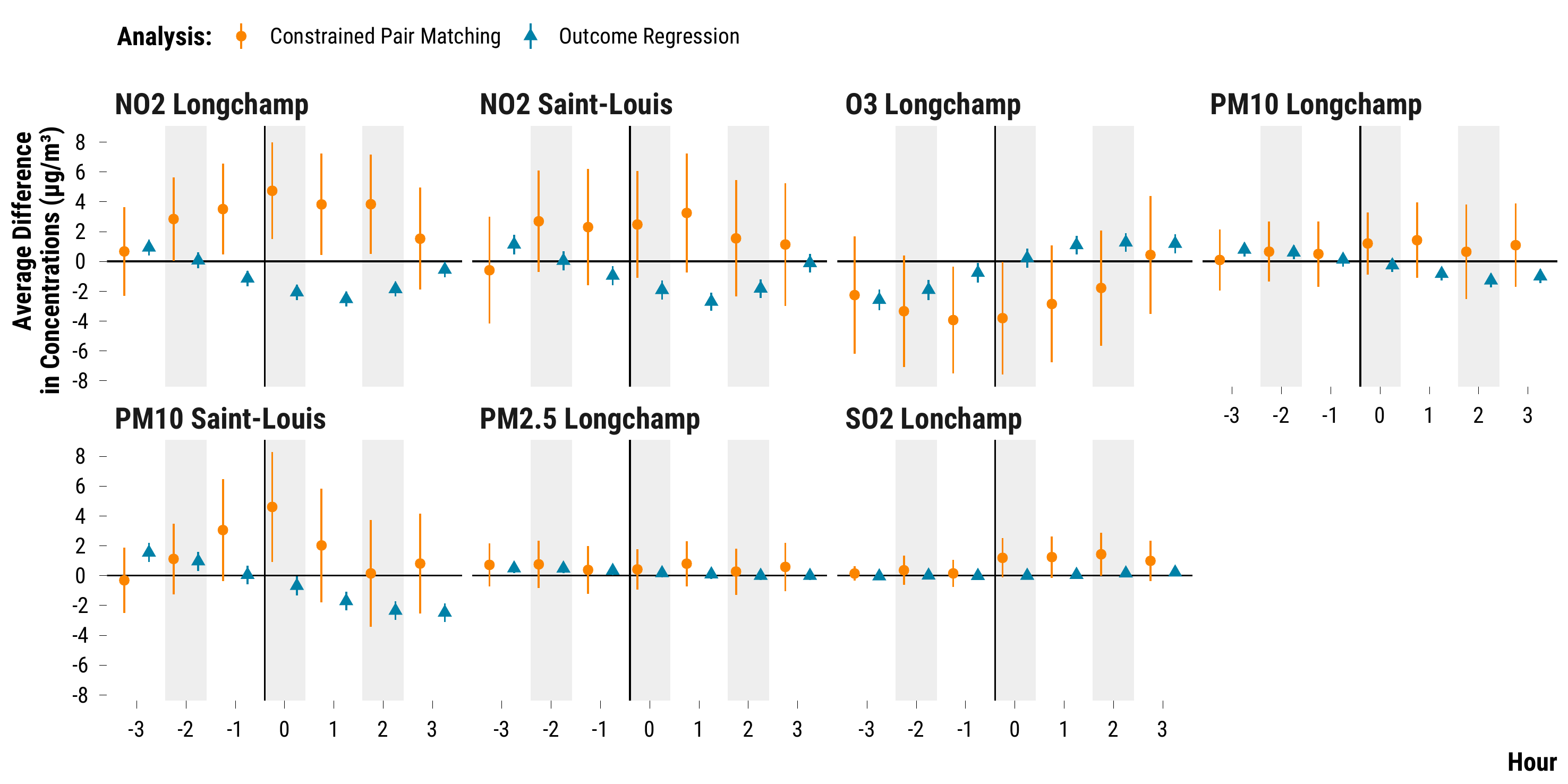}
\label{fig:graph_matching_regression}
\caption*{\textit{Notes:} The treatment occurs at hour 0. The orange color represents the results of our matching procedure while the blue color represents results from the regression approach. Orange dots and blue triangles represent the point estimate of the average treatment effect on a pollutant concentration. Lines are 95\% confidence intervals of average treatment effects consistent with the data. The effects are plotted from the third lag to the third lead.}
\end{figure}

At the \href{https://lzabrocki.github.io/cruise_air_pollution/2_9_daily_regression_analysis.html}{daily} level, estimates found with the regression approach are relatively similar but much more precise to those obtained with our matching procedure. However, at the \href{https://lzabrocki.github.io/cruise_air_pollution/1_9_hourly_regression_analysis.html}{hourly} level, regression estimates are of smaller magnitudes and even of opposite signs for some air pollutants (see \Cref{fig:graph_matching_regression}). This could be due to the fact that we are comparing two different samples. The alternative reason to explain this discrepancy could be due the multivariate regression model failure to correctly adjust for the functional forms of confounders and to its inherent tendency to extrapolate treatment effects outside the support of the data. Hourly results on the impact of vessel emission on air pollution are also more consistent with what has been observed in previous observational studies on the impact of cruise traffic on air pollutant concentrations \citep{diesch2013investigation, eckhardt2013influence, merico2016influence}.

\pagebreak




\section{Discussion}\label{sec:discussion}

In this section, we start by discussing our results in view of the environmental science literature. We then reflect on the new statistical methods used for our analyses. Finally, we suggest paths for future research assessing the causal impact of maritime traffic on air pollution when natural or policy experiments are not available.

\subsection{Putting our Results into Perspective}

Our results point to a potential short-term effect of cruise traffic on the concentrations of NO$_{2}$, O$_{3}$, SO$_{2}$, and PM$_{10}$ at the hourly level. At the daily level, we do not observe an impact of cruise vessel on all air pollutant concentrations. However, for both experiments our 95\% Fisherian intervals are often wide, and the implied degree of randomization-based uncertainty can be quite large relative to the average concentration of these air pollutants.

Directly comparing our results to those found in the atmospheric science literature is difficult for several reasons. First, they are based on other methods---either source apportionment techniques or dispersion modeling---and usually only report average effects without comparable measures of uncertainty. Second, they often consider the entire traffic of vessels rather than isolating the impact of a pre-defined treatment, as we do. Third, recent literature reviews have shown that the contribution of vessel emissions to local air pollution depends highly on the port-city considered and the procedure carried out by researchers \citep{viana2014impact, murena2018impact}. We can nonetheless assess whether our causal estimates are of the same order of magnitude as estimates from the atmospheric sciences literature.

For gaseous pollutants such as NO$_{2}$ and SO$_{2}$, the atmospheric science literature has mostly used emissions inventories combined with dispersion modeling \citep{viana2014impact}. The few studies on ports from the Mediterranean area find different contributions of maritime traffic to city-level concentrations depending on the size of the city, the location of the monitoring stations, the prevailing wind patterns, the type of boat considered and the assumptions used in the emissions inventory \citep{murena2018impact, mocerino2020methodology}. These estimates typically take into account all the phases where a vessel may contribute to pollution, in particular the hotelling phase, while we do not have information of the duration of the different phases. For NO$_{2}$, estimates range from 1.2-3.5\% for the contribution of cruise ships in summer in Naples \citep{murena2018impact} to 32.5\% for the contribution of all types of ships in the Italian city of Brindisi \citep{merico2017atmospheric}. Our estimated contribution of cruise traffic to NO$_{2}$ concentrations at the hourly level is equal to an increase of 16\%. The estimates for SO$_{2}$ range from 1.5\% for Naples in winter \citep{murena2018impact} to 46\% for Brindisi in summer \citep{merico2017atmospheric}. At the hourly level, we observe an increase of 52\% in SO$_{2}$ concentrations.

For particulate matter, source apportionment methods are commonly used \citep{sorte2020impact, viana2008source}. Estimates for the contribution of vessels to PM$_{10}$ concentrations range from 1.1\% for Rijeka in Croatia up to 11\% for Genoa in Italy \citep{merico2016influence, bove2014integrated}. We do not observe an effect on particulate matter in our daily experiment. This is however consistent with a measurement campaign carried out by Marseille's air quality monitoring agency \citep{atmosud_quelle_2019}, which relies on a five-month data collection near the port.

The media and non-governmental organizations have insisted on the high contribution of vessel traffic, and in particular cruise vessel traffic, to city-level emissions as measured by emission inventories. Our hourly experiment confirms that cruise traffic can increase air pollutant concentration on relatively short-time scale. Yet, the results of our daily experiment fail to suggest an impact of cruise vessels to air pollution on a longer time scale. We can contrast the results of our daily experiment on NO$_{2}$ concentrations with the contribution of road traffic, which can be inferred from a simple comparison between weekdays and weekends (see \href{https://lzabrocki.github.io/cruise_air_pollution/2_10_daily_road_traffic_analysis.html}{our simple road traffic analysis}). Because they are balanced in terms of weather covariates, the difference in observed concentrations between weekdays and weekends can be attributed to differences in economic activity only, and in particular to differences in road traffic. Road traffic decreases by 20\% on average on Saturdays and Sundays. In parallel, NO$_{2}$ concentrations decrease by 20\% of their average level at the Saint-Louis station. Although other sources of pollution may be less intense on weekends, the road traffic and NO$_{2}$ time series follow an extremely similar pattern, suggesting a strong contribution of road traffic to ambient concentrations compared to maritime traffic. Besides, cruise traffic tends to be higher on week-end and this positive flow of vessels does not offset the likely effect of road traffic on NO$_{2}$ concentrations. Beyond emission inventories informing on the relative contribution of different sectors to emissions, more systematic assessments based on observational studies are needed to understand the relative contribution of different sources to ambient concentrations. It would help better evaluate the benefits of abatement in each sector and prioritize policies. Besides, the higher salience of plumes emitted by cruise vessels and the potential larger concerns over this source of air pollution is an interesting area for future research. 

\subsection{Reflection on the Methods}

The causal inference pipeline we follow helps to clearly distinguish the design stage of our study---where we create hypothetical experiments---from its statistical analysis. Our pair-matching procedure has two notable advantages. First, it prunes treated units for which we cannot find a similar control unit, and thereby avoids extrapolating treatment effects for units without any empirical counterfactuals. In a way, a matching procedure reveals the common support available in the data from which we can draw our statistical inference upon. Second, our approach adjusts for covariates in a nonparametric way and achieves balance between treated and control units on observed covariates. This is another advantage, as it is often hard to guess what functional forms are needed to adjust for confounding factors \citep{cochran1973controlling, ho2007matching, imbens2015matching}. 

Yet, matching applied to high-frequency and regular vessel traffic data also poses difficulties. First, finding comparable treated and control units is challenging. At the hourly level, it is difficult to match a treated unit with a control unit because vessel traffic is very regular within different periods of the year. For instance, cruise vessels nearly always dock in the port at particular hours and days of the week---leaving few control hours without any cruise traffic. Second, obtaining days with close weather patterns over several consecutive days is extremely difficult: at the hourly level, it was nearly impossible to find similar pairs over three lags of covariates. Surprisingly, even if we strive to find similar pairs of treated and control units, we observe a wide heterogeneity in pair differences in pollutant concentrations, which makes it difficult to precisely estimate the potential contribution of vessel emissions. In our study, we are therefore confronted with a trade-off between the comparability of units within pairs and the sample size on which we base our statistical analysis. 

Regarding the statistical inference procedure, randomization-based inference allows us to avoid large-sample approximations and makes no assumption on the distribution of our test statistic under the sharp null hypothesis \citep{rosenbaum2010design}. Given that we deal with small sample sizes, we believe that our procedure is a relevant alternative to standard quantification of uncertainty. Yet, randomization-based inference procedure relies on the stringent assumption that the treatment is constant. This is arguably an unrealistic assumption. We therefore provide results from a Neymanian inference perspective \citep{splawa1990application, imbens2015causal}, which considers average treatment effects rather than unit-level treatment effects. Although based on a different interpretation of the data, results from Fisherian and Neymanian inference are very similar. The recent approach proposed by \cite{wu2021randomization} to make a randomization inference procedure conservative for average treatment effects also give very similar results. As an alternative to Fisherian and Neyman modes of inference, we could also have implemented a Bayesian model-based approach, which explicitly imputes the missing potential outcomes given the observed data and can target a larger variety of causal estimands \citep{rubin1978bayesian, imbens2015causal, bind2019bridging}.

Finally, in terms of feasibility and data requirement, our approach could in some contexts be less costly to implement than source apportionment or dispersion modeling from the atmospheric sciences: while emission inventories and monitoring campaigns are commonplace in developed countries, they are less common in developing country contexts. In contrast, our method relies on vessel traffic data that is collected by all port authorities as part of their business activities, and can also be accessed via online platform.  Pollution data from monitoring stations are often publicly available too, and could be replaced by satellite data in contexts where monitoring stations are sparse.

\subsection{Potential Paths for Future Research}

We see at least three main improvements for future research aiming to estimate the effects of maritime traffic on air pollution. First, it would be useful to exploit data on the duration vessels keep their engines running while docked at the port. Several studies indicate that a large share of air pollutant emissions occur during this phase \citep{caiman2015, murena2018impact}. Second, monitoring stations in Marseille only measure some air pollutants and are located relatively far away from the port. It would be useful to carry out similar analyses as ours in a port city where pollutants such as ultrafine particles are monitored and with receptors located in the port at different heights \citep{viana2014impact, mocerino2020methodology}. Besides, the weather data we exploit are located 25km away from the city, which adds noise. It would also be useful to have a monitoring station located within the city. Third, we saw that a non-negligible fraction of our matched pairs could be affected by spillover effects. New matching algorithms should be developed to force treated and control units from different pairs to be temporally far away: estimated treatment effects would be then less likely to be contaminated by spillovers.

We provide detailed replication materials in the hope that researchers could implement but also improve our method. Even if there remain challenges with regards to potential spillover effects and the imprecision of estimates, we believe that our study provides a fruitful and principled approach to try to estimate the impacts of maritime traffic on local air pollution.




\setlength\bibsep{0pt}

\bibliographystyle{econ}
\bibliography{bibliography.bib}

@Article{corbett2007mortality,
  author    = {Corbett, James J. and Winebrake, James J. and Green, Erin H. and Kasibhatla, Prasad and Eyring, Veronika and Lauer, Axel},
  journal   = {Environmental science \& technology},
  title     = {Mortality from ship emissions: a global assessment},
  year      = {2007},
  number    = {24},
  pages     = {8512--8518},
  volume    = {41},
  publisher = {ACS Publications},
}

@Article{sorte2020impact,
  author    = {Sorte, Sandra and Rodrigues, Vera and Borrego, Carlos and Monteiro, Alexandra},
  journal   = {Environmental Pollution},
  title     = {Impact of harbour activities on local air quality: A review},
  year      = {2020},
  pages     = {113542},
  volume    = {257},
  publisher = {Elsevier},
}

@Article{contini2011direct,
  author    = {Contini, D and Gambaro, A and Belosi, F and De Pieri, S and Cairns, WRL and Donateo, A and Zanotto, E and Citron, M},
  journal   = {Journal of Environmental Management},
  title     = {The direct influence of ship traffic on atmospheric {PM2. 5}, {PM10} and {PAH} in {V}enice},
  year      = {2011},
  number    = {9},
  pages     = {2119--2129},
  volume    = {92},
  publisher = {Elsevier},
}

@Article{moretti2011pollution,
  author    = {Moretti, Enrico and Neidell, Matthew},
  journal   = {Journal of Human Resources},
  title     = {Pollution, health, and avoidance behavior evidence from the ports of {L}os {A}ngeles},
  year      = {2011},
  number    = {1},
  pages     = {154--175},
  volume    = {46},
  publisher = {University of Wisconsin Press},
}

@Article{rubin1974estimating,
  author    = {Rubin, Donald B},
  journal   = {Journal of Educational Psychology},
  title     = {Estimating causal effects of treatments in randomized and nonrandomized studies.},
  year      = {1974},
  number    = {5},
  pages     = {688},
  volume    = {66},
  publisher = {American Psychological Association},
}

@Article{holland1986statistics,
  author    = {Holland, Paul W},
  journal   = {Journal of the American statistical Association},
  title     = {Statistics and causal inference},
  year      = {1986},
  number    = {396},
  pages     = {945--960},
  volume    = {81},
  publisher = {Taylor \& Francis},
}

@Article{rubin2005causal,
  author    = {Rubin, Donald B},
  journal   = {Journal of the American Statistical Association},
  title     = {Causal inference using potential outcomes: Design, modeling, decisions},
  year      = {2005},
  number    = {469},
  pages     = {322--331},
  volume    = {100},
  publisher = {Taylor \& Francis},
}

@Article{sommer2018comparing,
  author    = {Sommer, Alice J and Lee, Mihye and Bind, Marie-Ab{\`e}le C},
  journal   = {Palgrave communications},
  title     = {Comparing apples to apples: an environmental criminology analysis of the effects of heat and rain on violent crimes in {B}oston},
  year      = {2018},
  number    = {1},
  pages     = {1--10},
  volume    = {4},
  publisher = {Palgrave},
}

@Article{ho2007matching,
  author    = {Ho, Daniel E and Imai, Kosuke and King, Gary and Stuart, Elizabeth A},
  journal   = {Political analysis},
  title     = {Matching as nonparametric preprocessing for reducing model dependence in parametric causal inference},
  year      = {2007},
  number    = {3},
  pages     = {199--236},
  volume    = {15},
  publisher = {Cambridge University Press},
}

@Article{sommer2021assessing,
  author    = {Sommer, Alice J and Leray, Emmanuelle and Lee, Young and Bind, Marie-Ab{\`e}le C},
  journal   = {Statistics in Medicine},
  title     = {Assessing environmental epidemiology questions in practice with a causal inference pipeline: An investigation of the air pollution-multiple sclerosis relapses relationship},
  year      = {2021},
  number    = {6},
  pages     = {1321--1335},
  volume    = {40},
  publisher = {Wiley Online Library},
}

@Article{fisher1937design,
  author    = {Fisher, Ronald Aylmer and others},
  journal   = {The design of experiments.},
  title     = {The design of experiments.},
  year      = {1937},
  publisher = {Oliver \& Boyd, Edinburgh \& London.},
}

@Book{rosenbaum2010design,
  author    = {Rosenbaum, Paul R and others},
  publisher = {Springer},
  title     = {Design of observational studies},
  year      = {2010},
  volume    = {10},
}

@Article{viana2014impact,
  author    = {Viana, Mar and Hammingh, Pieter and Colette, Augustin and Querol, Xavier and Degraeuwe, Bart and de Vlieger, Ina and van Aardenne, John},
  journal   = {Atmospheric Environment},
  title     = {Impact of maritime transport emissions on coastal air quality in {E}urope},
  year      = {2014},
  pages     = {96--105},
  volume    = {90},
  publisher = {Elsevier},
}

@Article{mueller2011ships,
  author    = {Mueller, Daniel and Uibel, Stefanie and Takemura, Masaya and Klingelhoefer, Doris and Groneberg, David A},
  journal   = {Journal of Occupational Medicine and Toxicology},
  title     = {Ships, ports and particulate air pollution-an analysis of recent studies},
  year      = {2011},
  number    = {1},
  pages     = {1--6},
  volume    = {6},
  publisher = {BioMed Central},
}

@Manual{missranger,
  title  = {missRanger: Fast Imputation of Missing Values},
  author = {Michael Mayer},
  note   = {R package version 2.1.0},
  year   = {2019},
  url    = {https://cran.r-project.org/package=missRanger},
}

@Article{bind2019bridging,
  author    = {Bind, Marie-Abèle C. and Rubin, Donald B.},
  journal   = {Statistical Methods in Medical Research},
  title     = {Bridging observational studies and randomized experiments by embedding the former in the latter},
  year      = {2019},
  number    = {7},
  pages     = {1958--1978},
  volume    = {28},
  publisher = {SAGE Publications Sage UK: London, England},
}

@Article{rubin1991practical,
  author    = {Rubin, Donald B},
  journal   = {Biometrics},
  title     = {Practical implications of modes of statistical inference for causal effects and the critical role of the assignment mechanism},
  year      = {1991},
  pages     = {1213--1234},
  publisher = {JSTOR},
}

@Book{rubin2006matched,
  author    = {Rubin, Donald B},
  publisher = {Cambridge University Press},
  title     = {Matched sampling for causal effects},
  year      = {2006},
}

@Article{stuart2010matching,
  author    = {Stuart, Elizabeth A},
  journal   = {Statistical science: a review journal of the Institute of Mathematical Statistics},
  title     = {Matching methods for causal inference: A review and a look forward},
  year      = {2010},
  number    = {1},
  pages     = {1},
  volume    = {25},
  publisher = {NIH Public Access},
}

@Article{imbens2015matching,
  author    = {Imbens, Guido W},
  journal   = {Journal of Human Resources},
  title     = {Matching methods in practice: Three examples},
  year      = {2015},
  number    = {2},
  pages     = {373--419},
  volume    = {50},
  publisher = {University of Wisconsin Press},
}

@InProceedings{micali1980v,
  author       = {Micali, Silvio and Vazirani, Vijay V},
  booktitle    = {21st Annual Symposium on Foundations of Computer Science (sfcs 1980)},
  title        = {An O (v$\vert$ v$\vert$ c$\vert$ E$\vert$) algorithm for finding maximum matching in general graphs},
  year         = {1980},
  organization = {IEEE},
  pages        = {17--27},
}

@Book{imbens2015causal,
  author    = {Imbens, Guido W and Rubin, Donald B},
  publisher = {Cambridge University Press},
  title     = {Causal inference in statistics, social, and biomedical sciences},
  year      = {2015},
}

@Book{dasgupta2015,
  author    = {Dasgupta, Tirthankar and Rubin, Donald B},
  publisher = {Harvard university},
  title     = {STAT 240: Matched Sampling and Study Design},
  year      = {Fall 2015},
}

@Article{keele2012strengthening,
  author    = {Keele, Luke and McConnaughy, Corrine and White, Ismail},
  journal   = {American Journal of Political Science},
  title     = {Strengthening the experimenter’s toolbox: Statistical estimation of internal validity},
  year      = {2012},
  number    = {2},
  pages     = {484--499},
  volume    = {56},
  publisher = {Wiley Online Library},
}

@Article{sekhon2009opiates,
  author    = {Sekhon, Jasjeet S},
  journal   = {Annual Review of Political Science},
  title     = {Opiates for the matches: Matching methods for causal inference},
  year      = {2009},
  pages     = {487--508},
  volume    = {12},
  publisher = {Annual Reviews},
}

@Article{murena2018impact,
  author    = {Murena, F and Mocerino, L and Quaranta, F and Toscano, D},
  journal   = {Atmospheric Environment},
  title     = {Impact on air quality of cruise ship emissions in {N}aples, {I}taly},
  year      = {2018},
  pages     = {70--83},
  volume    = {187},
  publisher = {Elsevier},
}

@Article{merico2017atmospheric,
  author    = {Merico, Eva and Gambaro, Andrea and Argiriou, A and Alebic-Juretic, A and Barbaro, Elena and Cesari, Daniela and Chasapidis, Leonidas and Dimopoulos, S and Dinoi, Adelaide and Donateo, Antonio and others},
  journal   = {Transportation Research Part D: Transport and Environment},
  title     = {Atmospheric impact of ship traffic in four Adriatic-Ionian port-cities: Comparison and harmonization of different approaches},
  year      = {2017},
  pages     = {431--445},
  volume    = {50},
  publisher = {Elsevier},
}

@Article{viana2008source,
  author    = {Viana, Mar and Kuhlbusch, Thomas AJ and Querol, Xavier and Alastuey, Andr{\'e}s and Harrison, Roy M and Hopke, Philip K and Winiwarter, Wilfried and Vallius, M and Szidat, S{\"o}nke and Pr{\'e}v{\^o}t, Andr{\'e} SH and others},
  journal   = {Journal of Aerosol Science},
  title     = {Source apportionment of particulate matter in {E}urope: a review of methods and results},
  year      = {2008},
  number    = {10},
  pages     = {827--849},
  volume    = {39},
  publisher = {Elsevier},
}

@Article{merico2016influence,
  author    = {Merico, E and Donateo, A and Gambaro, A and Cesari, D and Gregoris, E and Barbaro, E and Dinoi, A and Giovanelli, G and Masieri, S and Contini, D},
  journal   = {Atmospheric Environment},
  title     = {Influence of in-port ships emissions to gaseous atmospheric pollutants and to particulate matter of different sizes in a {M}editerranean harbour in {I}taly},
  year      = {2016},
  pages     = {1--10},
  volume    = {139},
  publisher = {Elsevier},
}

@Article{bove2014integrated,
  author    = {Bove, MC and Brotto, P and Cassola, F and Cuccia, E and Massab{\`o}, D and Mazzino, A and Piazzalunga, A and Prati, P},
  journal   = {Atmospheric Environment},
  title     = {An integrated {PM2.5} source apportionment study: {P}ositive {M}atrix {F}actorisation vs. the chemical transport model {CAM}x},
  year      = {2014},
  pages     = {274--286},
  volume    = {94},
  publisher = {Elsevier},
}

@Article{cochran1973controlling,
  author    = {Cochran, William G. and Rubin, Donald B.},
  journal   = {Sankhy{\=a}: The Indian Journal of Statistics, Series A},
  title     = {Controlling bias in observational studies: A review},
  year      = {1973},
  pages     = {417--446},
  publisher = {JSTOR},
}

@Article{gelman2014beyond,
  author    = {Gelman, Andrew and Carlin, John},
  journal   = {Perspectives on Psychological Science},
  title     = {Beyond power calculations: Assessing type {S} (sign) and type {M} (magnitude) errors},
  year      = {2014},
  number    = {6},
  pages     = {641--651},
  volume    = {9},
  publisher = {Sage Publications Sage CA: Los Angeles, CA},
}

@Book{gelman2020regression,
  author    = {Gelman, Andrew and Hill, Jennifer and Vehtari, Aki},
  publisher = {Cambridge University Press},
  title     = {Regression and other stories},
  year      = {2020},
}

@Article{splawa1990application,
  author    = {Splawa-Neyman, Jerzy and Dabrowska, Dorota M and Speed, TP},
  journal   = {Statistical Science},
  title     = {On the application of probability theory to agricultural experiments. {E}ssay on principles. {S}ection 9.},
  year      = {1990},
  pages     = {465--472},
  publisher = {JSTOR},
}

@Article{mocerino2020methodology,
  author    = {Mocerino, Luigia and Murena, Fabio and Quaranta, Franco and Toscano, Domenico},
  journal   = {Scientific Reports},
  title     = {A methodology for the design of an effective air quality monitoring network in port areas},
  year      = {2020},
  number    = {1},
  pages     = {1--10},
  volume    = {10},
  publisher = {Nature Publishing Group},
}

@Article{zhu2021effects,
  author    = {Zhu, Junming and Wang, Jiali},
  journal   = {Journal of Environmental Economics and Management},
  title     = {The effects of fuel content regulation at ports on regional pollution and shipping industry},
  year      = {2021},
  pages     = {102424},
  volume    = {106},
  publisher = {Elsevier},
}

@Article{Friedrich2017,
  author  = {Friedrich, Axel},
  journal = {The Guardian},
  title   = {Heading to {V}enice? {D}on’t forget your pollution mask},
  year = {2017},
  url     = {https://www.theguardian.com/environment/2017/jul/31/heading-to-venice-dont-forget-your-pollution-mask},
}

@Article{Chrisafis2018,
  author  = {Chrisafis, Angelique},
  journal = {The Guardian},
  title   = {'{I} don't want ships to kill me': Marseille fights cruise liner pollution},
  year  =   {2018},
  url     = {https://www.theguardian.com/environment/2017/jul/31/heading-to-venice-dont-forget-your-pollution-mask},
}

@TechReport{caiman2015,
  author      = {CAIMAN},
  institution = {Technical Report},
  title       = {Air quality impact and greenhouse gases assessment for cruise and passenger ships},
  year        = {2015},

}

@TechReport{apice2013,
  author      = {Piga, Damien and Armengaud, Alexandre and Devèze, Magali and Parra, Michael and Marchand, Nicolas and Detournay, Anaïs and Dalia, Salameh},
  institution = {Technical Report},
  title       = {Synthèse du projet APICE - Marseille},
  year        = {2013}
}

@TechReport{atmosudbis2018,
  author      = {AtmoSud},
  institution = {Technical Report},
  title       = {Quelle qualité de l’air pour les riverains des ports de Nice et Marseille?},
  year        = {2018}
}

@Article{rubin1978bayesian,
  author    = {Rubin, Donald B},
  journal   = {The Annals of Statistics},
  title     = {Bayesian inference for causal effects: The role of randomization},
  year      = {1978},
  pages     = {34--58},
  publisher = {JSTOR},
}

@book{dasgupta_rubin_2021,
	title = {Experimental Design: A Randomization-Based Perspective},
	publisher = {Unpublished Textbook},
	author = {Dasgupta, Tirthankar and Rubin, Donald B.},
	year = {2021}
}

@article{zhong2017traffic,
  title={Traffic congestion, ambient air pollution, and health: Evidence from driving restrictions in Beijing},
  author={Zhong, Nan and Cao, Jing and Wang, Yuzhu},
  journal={Journal of the Association of Environmental and Resource Economists},
  volume={4},
  number={3},
  pages={821--856},
  year={2017},
  publisher={University of Chicago Press Chicago, IL}
}

@article{giaccherini2021particulate,
  title={When particulate matter strikes cities: Social disparities and health costs of air pollution},
  author={Giaccherini, Matilde and Kopinska, Joanna and Palma, Alessandro},
  journal={Journal of Health Economics},
  volume={78},
  pages={102478},
  year={2021},
  publisher={Elsevier}
}

@techreport{godzinski2019short,
  title={Short-term health effects of public transport disruptions: air pollution and viral spread channels},
  author={Godzinski, Alexandre and Castillo, M Suarez and others},
  year={2019},
  institution={Institut National de la Statistique et des Etudes Economiques}
}

@article{schlenker2016airports,
  title={Airports, air pollution, and contemporaneous health},
  author={Schlenker, Wolfram and Walker, W Reed},
  journal={The Review of Economic Studies},
  volume={83},
  number={2},
  pages={768--809},
  year={2016},
  publisher={Oxford University Press}
}

@article{knittel2016caution,
  title={Caution, drivers! Children present: Traffic, pollution, and infant health},
  author={Knittel, Christopher R and Miller, Douglas L and Sanders, Nicholas J},
  journal={Review of Economics and Statistics},
  volume={98},
  number={2},
  pages={350--366},
  year={2016},
  publisher={The MIT Press}
}

@article{bauernschuster2017labor,
  title={When labor disputes bring cities to a standstill: The impact of public transit strikes on traffic, accidents, air pollution, and health},
  author={Bauernschuster, Stefan and Hener, Timo and Rainer, Helmut},
  journal={American Economic Journal: Economic Policy},
  volume={9},
  number={1},
  pages={1--37},
  year={2017}
}

@article{simeonova2021congestion,
  title={Congestion pricing, air pollution, and children’s health},
  author={Simeonova, Emilia and Currie, Janet and Nilsson, Peter and Walker, Reed},
  journal={Journal of Human Resources},
  volume={56},
  number={4},
  pages={971--996},
  year={2021},
  publisher={University of Wisconsin Press}
}

@book{rosenbaum2018observation,
  title={Observation and experiment},
  author={Rosenbaum, Paul},
  year={2018},
  publisher={Harvard University Press}
}

@book{gerber2012field,
  title={Field experiments: Design, analysis, and interpretation},
  author={Gerber, Alan S and Green, Donald P},
  year={2012},
  publisher={WW Norton}
}

@article{cohen2010free,
  title={Free distribution or cost-sharing? Evidence from a randomized malaria prevention experiment},
  author={Cohen, Jessica and Dupas, Pascaline},
  journal={The Quarterly Journal of Economics},
  pages={1--45},
  year={2010},
  publisher={JSTOR}
}

@article{bowers2011fisher,
  title={Fisher's randomizationmode of statistical inference, then and now.},
  author={Bowers, Jake and Panagopoulos, Costas},
  journal = {Working Paper},
  year={2011},
  publisher={Citeseer}
}

@incollection{bowers2020causality,
  title={Causality and Design-Based Inference},
  author={Bowers, Jake and Leavitt, Thomas},
  booktitle={The SAGE Handbook of Research Methods in Political Science and International Relations},
  pages={769--804},
  year={2020},
  publisher={SAGE Publications Ltd}
}

@article{ho2006randomization,
  title={Randomization inference with natural experiments: An analysis of ballot effects in the 2003 California recall election},
  author={Ho, Daniel E and Imai, Kosuke},
  journal={Journal of the American statistical association},
  volume={101},
  number={475},
  pages={888--900},
  year={2006},
  publisher={Taylor \& Francis}
}

@article{hess2017randomization,
  title={Randomization inference with Stata: A guide and software},
  author={He{\ss}, Simon},
  journal={The Stata Journal},
  volume={17},
  number={3},
  pages={630--651},
  year={2017},
  publisher={SAGE Publications Sage CA: Los Angeles, CA}
}

@incollection{athey2017econometrics,
  title={The econometrics of randomized experiments},
  author={Athey, Susan and Imbens, Guido W},
  booktitle={Handbook of economic field experiments},
  volume={1},
  pages={73--140},
  year={2017},
  publisher={Elsevier}
}

@article{zhao2021covariate,
  title={Covariate-adjusted Fisher randomization tests for the average treatment effect},
  author={Zhao, Anqi and Ding, Peng},
  journal={Journal of Econometrics},
  volume={225},
  number={2},
  pages={278--294},
  year={2021},
  publisher={Elsevier}
}

@article{wu2021randomization,
  title={Randomization tests for weak null hypotheses in randomized experiments},
  author={Wu, Jason and Ding, Peng},
  journal={Journal of the American Statistical Association},
  volume={116},
  number={536},
  pages={1898--1913},
  year={2021},
  publisher={Taylor \& Francis}
}

@article{ding2016randomization,
  title={Randomization inference for treatment effect variation},
  author={Ding, Peng and Feller, Avi and Miratrix, Luke},
  journal={Journal of the Royal Statistical Society: Series B (Statistical Methodology)},
  volume={78},
  number={3},
  pages={655--671},
  year={2016},
  publisher={Wiley Online Library}
}

@article{caughey2021randomization,
  title={Randomization Inference beyond the Sharp Null: Bounded Null Hypotheses and Quantiles of Individual Treatment Effects},
  author={Caughey, Devin and Dafoe, Allan and Li, Xinran and Miratrix, Luke},
  journal={arXiv preprint arXiv:2101.09195},
  year={2021}
}

@article{keele2019randomization,
  title={Randomization inference for outcomes with clumping at zero},
  author={Keele, Luke and Miratrix, Luke},
  journal={The American Statistician},
  volume={73},
  number={2},
  pages={141--150},
  year={2019},
  publisher={Taylor \& Francis}
}

@article{cattaneo2015randomization,
  title={Randomization inference in the regression discontinuity design: An application to party advantages in the US Senate},
  author={Cattaneo, Matias D and Frandsen, Brigham R and Titiunik, Rocio},
  journal={Journal of Causal Inference},
  volume={3},
  number={1},
  pages={1--24},
  year={2015},
  publisher={De Gruyter}
}

@article{mackinnon2020randomization,
  title={Randomization inference for difference-in-differences with few treated clusters},
  author={MacKinnon, James G and Webb, Matthew D},
  journal={Journal of Econometrics},
  volume={218},
  number={2},
  pages={435--450},
  year={2020},
  publisher={Elsevier}
}

@article{baccini2017assessing,
  title={Assessing the short term impact of air pollution on mortality: a matching approach},
  author={Baccini, Michela and Mattei, Alessandra and Mealli, Fabrizia and Bertazzi, Pier Alberto and Carugno, Michele},
  journal={Environmental Health},
  volume={16},
  number={1},
  pages={1--12},
  year={2017},
  publisher={Springer}
}

@article{lee2021discovering,
  title={Discovering heterogeneous exposure effects using randomization inference in air pollution studies},
  author={Lee, Kwonsang and Small, Dylan S and Dominici, Francesca},
  journal={Journal of the American Statistical Association},
  volume={116},
  number={534},
  pages={569--580},
  year={2021},
  publisher={Taylor \& Francis}
}

@article{bind2019causal,
  title={Causal modeling in environmental health},
  author={Bind, Marie-Ab{\`e}le},
  journal={Annual review of public health},
  volume={40},
  pages={23--43},
  year={2019},
  publisher={Annual Reviews}
}

@article{zigler2014point,
  title={Point: clarifying policy evidence with potential-outcomes thinking—beyond exposure-response estimation in air pollution epidemiology},
  author={Zigler, Corwin Matthew and Dominici, Francesca},
  journal={American journal of epidemiology},
  volume={180},
  number={12},
  pages={1133--1140},
  year={2014},
  publisher={Oxford University Press}
}

@article{gutman2012analyses,
  title={Analyses that Inform Policy Decisions [with Discussions]},
  author={Gutman, R and Rubin, DB and Vansteelandt, Stijn},
  journal={Biometrics},
  volume={68},
  number={3},
  pages={671--678},
  year={2012},
  publisher={JSTOR}
}

@article{neyman1923applications,
  title={Sur les applications de la thar des probabilities aux experiences Agaricales: Essay des principle. Excerpts reprinted (1990) in English},
  author={Neyman, Jerzy},
  journal={Statistical Science},
  volume={5},
  number={463-472},
  pages={4},
  year={1923}
}

@article{forastiere2020assessing,
  title={Assessing short-term impact of PM 10 on mortality using a semiparametric generalized propensity score approach},
  author={Forastiere, Laura and Carugno, Michele and Baccini, Michela},
  journal={Environmental Health},
  volume={19},
  number={1},
  pages={1--13},
  year={2020},
  publisher={BioMed Central}
}

@article{liu_health_2016,
	title = {Health and climate impacts of ocean-going vessels in {East} {Asia}},
	volume = {6},
	copyright = {2016 Nature Publishing Group},
	issn = {1758-6798},
	url = {https://www.nature.com/articles/nclimate3083},
	doi = {10.1038/nclimate3083},
	language = {en},
	number = {11},
	journal = {Nature Climate Change},
	author = {Liu, Huan and Fu, Mingliang and Jin, Xinxin and Shang, Yi and Shindell, Drew and Faluvegi, Greg and Shindell, Cary and He, Kebin},
	year = {2016}
}

@article{liu_emissions_2019,
	title = {Emissions and health impacts from global shipping embodied in {US}–{China} bilateral trade},
	volume = {2},
	copyright = {2019 The Author(s), under exclusive licence to Springer Nature Limited},
	issn = {2398-9629},
	url = {https://www.nature.com/articles/s41893-019-0414-z},
	doi = {10.1038/s41893-019-0414-z},
	language = {en},
	number = {11},
	journal = {Nature Sustainability},
	author = {Liu, Huan and Meng, Zhi-Hang and Lv, Zhao-Feng and Wang, Xiao-Tong and Deng, Fan-Yuan and Liu, Yang and Zhang, Yan-Ni and Shi, Meng-Shuang and Zhang, Qiang and He, Ke-Bin},
	year = {2019}
}

@techreport{atmosud_quelle_2019,
	title = {Quelle qualité de l’air pour les riverains des ports de {Nice} et {Marseille}? Campagnes de mesure 2018},
	url = {https://www.atmosud.org/sites/paca/files/atoms/files/200511_synthese_travaux_ports_2018_0.pdf},
	author = {{Atmosud}},
	institution = {Atmosud},
	year = {2019}
}

@article{greifer2021matching,
  title={Matching methods for confounder adjustment: an addition to the epidemiologist’s toolbox},
  author={Greifer, Noah and Stuart, Elizabeth A},
  journal={Epidemiologic reviews},
  volume={43},
  number={1},
  pages={118--129},
  year={2021},
  publisher={Oxford University Press}
}

@article{fogarty2020studentized,
  title={Studentized sensitivity analysis for the sample average treatment effect in paired observational studies},
  author={Fogarty, Colin B},
  journal={Journal of the American Statistical Association},
  volume={115},
  number={531},
  pages={1518--1530},
  year={2020},
  publisher={Taylor \& Francis}
}

@article{rosenbaum1987sensitivity,
  title={Sensitivity analysis for certain permutation inferences in matched observational studies},
  author={Rosenbaum, Paul R},
  journal={Biometrika},
  volume={74},
  number={1},
  pages={13--26},
  year={1987},
  publisher={Oxford University Press}
}

@article{lu2019note,
  title={A note on Type S/M errors in hypothesis testing},
  author={Lu, Jiannan and Qiu, Yixuan and Deng, Alex},
  journal={British Journal of Mathematical and Statistical Psychology},
  volume={72},
  number={1},
  pages={1--17},
  year={2019},
  publisher={Wiley Online Library}
}

@article{branson2021randomization,
  title={Randomization Tests to Assess Covariate Balance When Designing and Analyzing Matched Datasets},
  author={Branson, Zach},
  journal={Observational Studies},
  volume={7},
  number={2},
  pages={1--36},
  year={2021},
  publisher={University of Pennsylvania Press}
}

@article{diesch2013investigation,
  title={Investigation of gaseous and particulate emissions from various marine vessel types measured on the banks of the Elbe in Northern Germany},
  author={Diesch, J-M and Drewnick, F and Klimach, T and Borrmann, S},
  journal={Atmospheric Chemistry and Physics},
  volume={13},
  number={7},
  pages={3603--3618},
  year={2013},
  publisher={Copernicus GmbH}
}

@article{eckhardt2013influence,
  title={The influence of cruise ship emissions on air pollution in Svalbard--a harbinger of a more polluted Arctic?},
  author={Eckhardt, Sabine and Hermansen, Ove and Grythe, Henrik and Fiebig, Markus and Stebel, Kerstin and Cassiani, Massimo and B{\"a}cklund, Are and Stohl, Andreas},
  journal={Atmospheric Chemistry and Physics},
  volume={13},
  number={16},
  pages={8401--8409},
  year={2013},
  publisher={Copernicus GmbH}
}

@article{king2006dangers,
  title={The dangers of extreme counterfactuals},
  author={King, Gary and Zeng, Langche},
  journal={Political analysis},
  volume={14},
  number={2},
  pages={131--159},
  year={2006},
  publisher={Cambridge University Press}
}

@article{tufte1985visual,
  title={The visual display of quantitative information},
  author={Tufte, Edward R},
  journal={The Journal for Healthcare Quality (JHQ)},
  volume={7},
  number={3},
  pages={15},
  year={1985},
  publisher={LWW}
}

@book{cleveland1993visualizing,
  title={Visualizing data},
  author={Cleveland, William S},
  year={1993},
  publisher={Hobart press}
}

@Manual{retrodesign,
    title = {retrodesign: Tools for Type S (Sign) and Type M (Magnitude) Errors},
    author = {Andrew Timm},
    year = {2019},
    note = {R package version 0.1.0},
    url = {https://CRAN.R-project.org/package=retrodesign},
}

@techreport{hansen2022uncharted,
  title={Uncharted Waters: Effects of Maritime Emission Regulation},
  author={Hansen-Lewis, Jamie and Marcus, Michelle M},
  year={2022},
  institution={National Bureau of Economic Research}
}

@book{tukey1977exploratory,
  title={Exploratory data analysis},
  author={Tukey, John W and others},
  volume={2},
  year={1977},
  publisher={Reading, MA}
}

@article{rubin2008objective,
  title={For objective causal inference, design trumps analysis},
  author={Rubin, Donald B},
  journal={The annals of applied statistics},
  volume={2},
  number={3},
  pages={808--840},
  year={2008},
  publisher={Institute of Mathematical Statistics}
}

@article{klotz2022local,
  title={Local Standards, Behavioral Adjustments, and Welfare: Evaluating California’s Ocean-Going Vessel Fuel Rule},
  author={Klotz, Richard and Berazneva, Julia},
  journal={Journal of the Association of Environmental and Resource Economists},
  volume={9},
  number={3},
  pages={383--424},
  year={2022},
  publisher={The University of Chicago Press Chicago, IL}
}

@article{abadie2020sampling,
  title={Sampling-Based versus Design-Based Uncertainty in Regression Analysis},
  author={Abadie, Alberto and Athey, Susan and Imbens, Guido W and Wooldridge, Jeffrey M},
  journal={Econometrica},
  volume={88},
  number={1},
  pages={265--296},
  year={2020},
  publisher={Wiley Online Library}
}

\pagebreak

\appendix

\section{Appendix}
\setcounter{figure}{0}
\renewcommand\thefigure{\thesection.\arabic{figure}}

\begin{figure}[!ht]
\centering
\caption{Effects of Cruise Vessel Traffic on Pollutant Concentrations at the Daily Level: Comparing Fisherian and Neymanian Modes of Inference.}
\includegraphics[width=\linewidth]{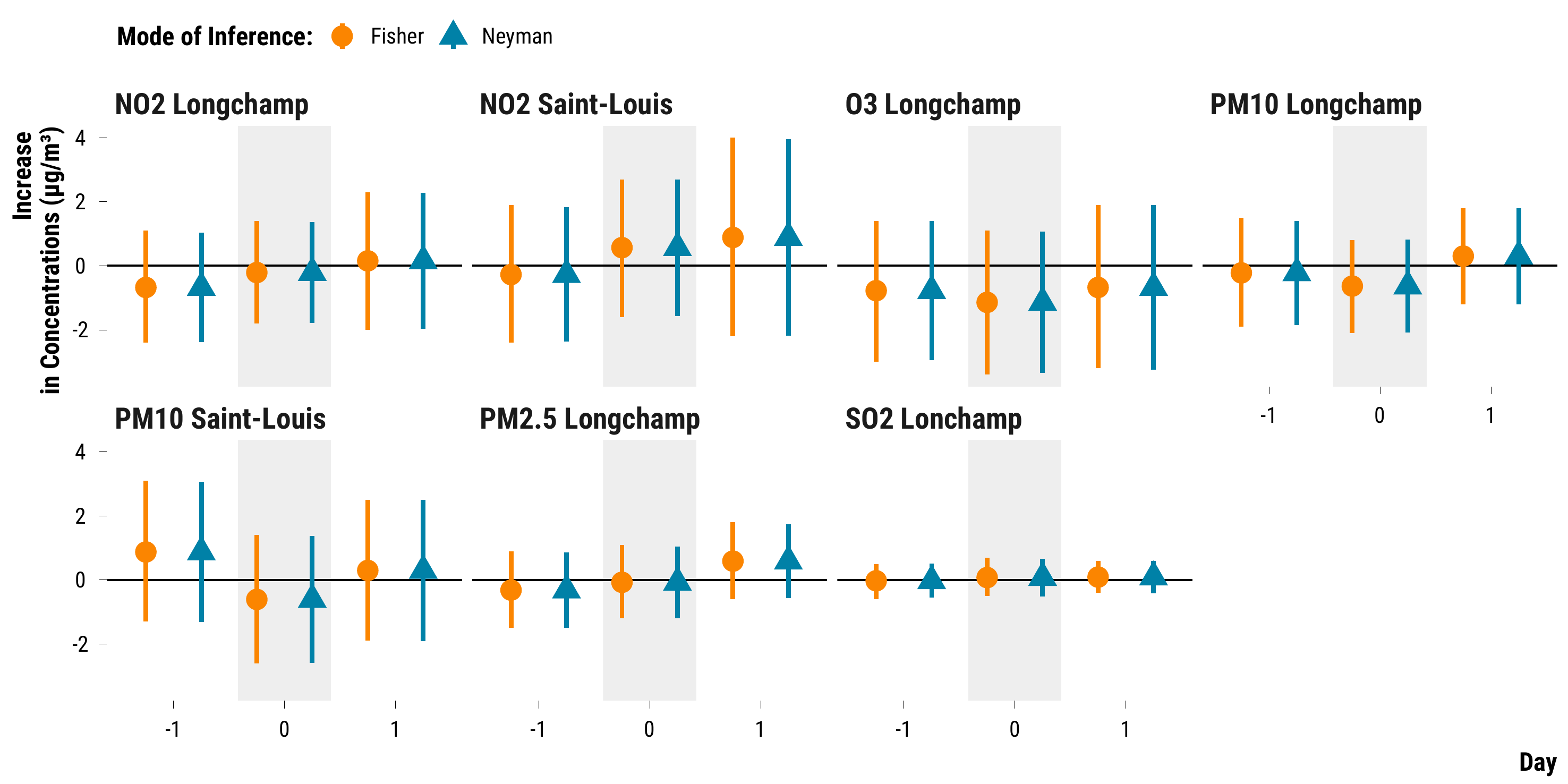}
\label{fig:graph_ci_neyman_daily}
\caption*{\textit{Notes:} The treatment occurs at day 0. Dots represent the point estimate effect on a pollutant concentration. Orange lines are 95\% Fisherian intervals of constant treatment effects consistent with the data. Blue lines are 95\% confidence intervals based on Neyman's mode of inference: they represent the range of average treatment effects consistent with the data. The effects are plotted from the first lag to the first lead.}
\end{figure}

\begin{figure}[!ht]
\centering
\caption{Effects of Cruise Vessel Traffic on Pollutant Concentrations at the Daily Level: Comparing Neyman's Mode of Inference to \cite{wu2021randomization}'s Randomization Inference Approach for Weak Null Hypotheses.}
\includegraphics[width=\linewidth]{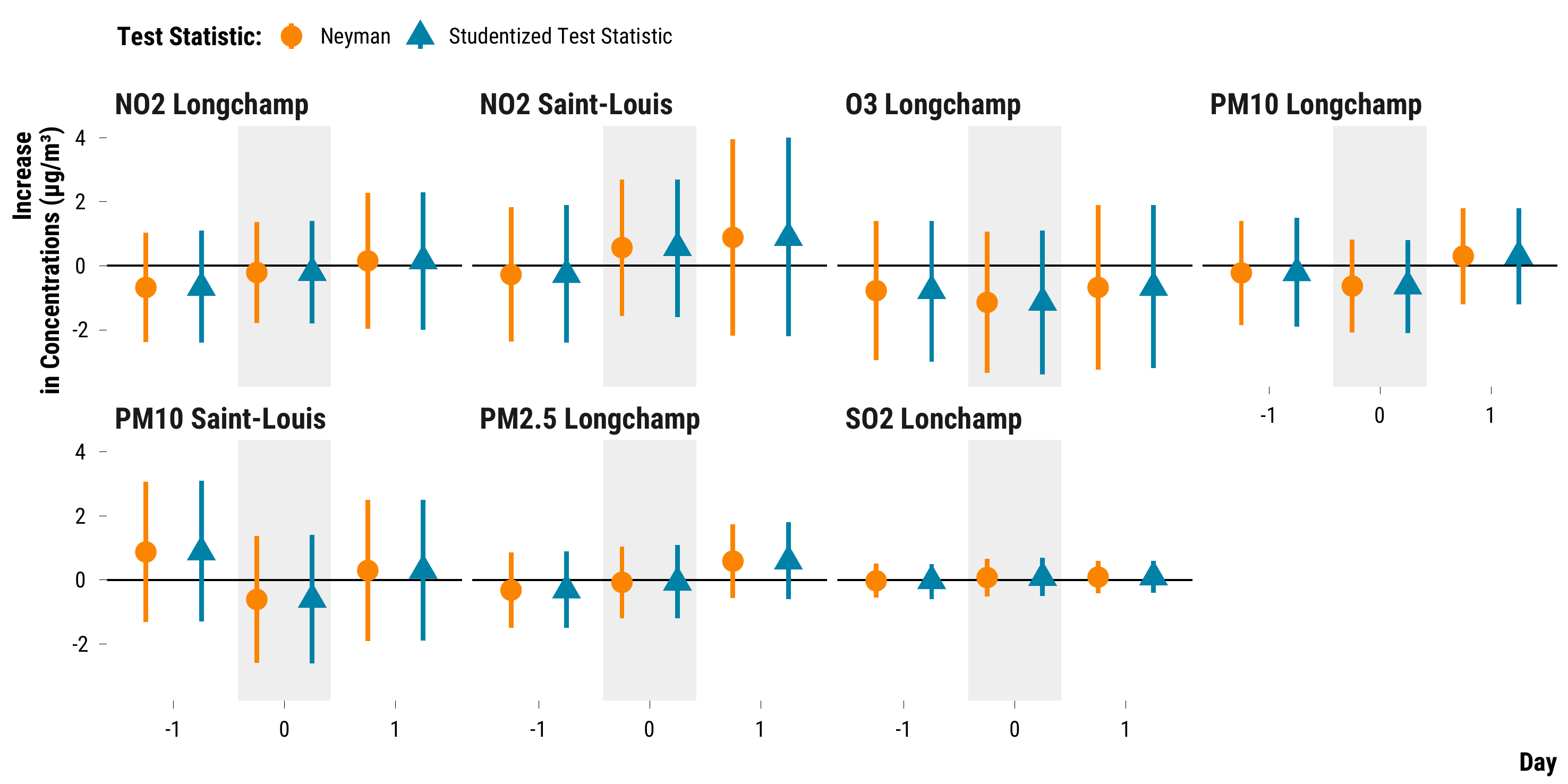}
\label{fig:graph_ci_fisher_weak_daily}
\caption*{\textit{Notes:} The treatment occurs at day 0. Dots represent the point estimate effect on a pollutant concentration. Orange lines are 95\% Neymanian intervals of average treatment effects consistent with the data. Blue lines are 95\% Fisherian intervals based on \cite{wu2021randomization}'s approach to make randomization inference asymptotically conservative for weak null hypotheses. The effects are plotted from the first lag to the first lead.}
\end{figure}

\begin{figure}[!ht]
\centering
\caption{Effects of Cruise Vessel Traffic on Pollutant Concentrations at the Hourly Level by Wind Directions.}
\includegraphics[width=\linewidth]{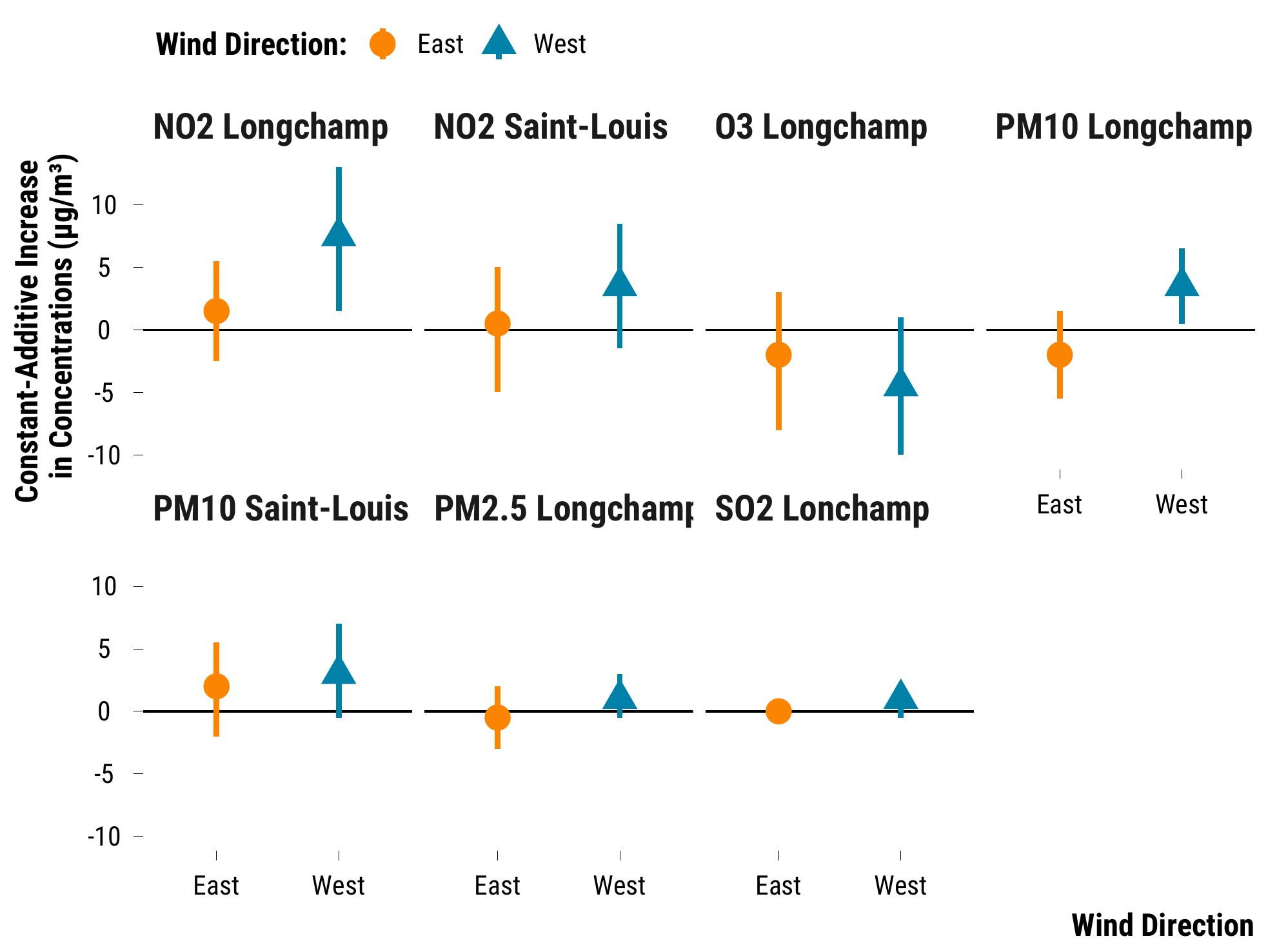}
\label{fig:graph_wd_hourly}
\caption*{\textit{Notes:} Dots represent the point estimate for the immediate effect of cruise traffic on hourly pollutant concentration. Lines are 95\% Fisherian intervals of constant treatment effects consistent with the data. Orange lines represent the effects for winds blowing from the East while orange lines represent effects for winds blowing from the West.}
\end{figure}

\end{document}